\documentclass[12pt]{article}
\usepackage{amsfonts,amsmath,epsf}
\advance\textheight by \topskip
\newcounter{defthm}
 
\newtheorem{defthm}{\whattheorem}[section]

\newcommand\sect[1]{\section{#1}\setcounter{equation}0\setcounter{defthm}0}

\newcommand\void[1]       {}

\newcommand\be            {\begin{equation}}
\newcommand\bea           {\begin{equation}\begin{array}l\displaystyle}
\newcommand\bearll        {\begin{array}{ll}\displaystyle}
\newcommand\ee            {\end{equation}}
\newcommand\eear          {\end{array}}
\newcommand\enl           {\\[1em]\displaystyle}
\newcommand\etb           {& \displaystyle}
\newcommand\erf[1] {(\ref{#1})}
\newcommand\labl[1]       {\label{#1}\ee}

\newcommand\bnd[1] {\raisebox{-.15em}{$\big|_{#1}$}}
\newcommand\Lo            {{\overline L}}
\newcommand\corr[1]       {\big\langle #1 \big\rangle}

\newcommand\eps           {\varepsilon}
\newcommand\id            {{\rm id}}

\newcommand\one           {{\bf1}}
\newcommand\orl           {\mathcal{O}_{\rm log}}

\newcommand\Cb            {\mathbb{C}}
\newcommand\Pb            {\mathbb{P}}
\newcommand\Rb            {\mathbb{R}}
\newcommand\Zb            {\mathbb{Z}}

\newcommand\Cc            {\mathcal{C}}
\newcommand\Nc            {\mathcal{N}}
\newcommand\Vc            {\mathcal{V}}

\newcommand\Dr            {\mathrm{D}}
\newcommand\Nr            {\mathrm{N}}

\newcommand\bmu           {{\boldsymbol{\mu}}}

\newcommand\brho          {{\boldsymbol{\rho}}}
\newcommand\bomega        {{\boldsymbol{\omega}}}
\newcommand\bOmega        {{\boldsymbol{\Omega}}}
\newcommand\bphi          {{\boldsymbol{\phi}}}
\newcommand\bpsi          {{\boldsymbol{\psi}}}

\newcommand{\om} {\omega}
\newcommand{\ol} {\overline}
\newcommand{\Om} {\Omega}
\newcommand{\Ut} {{\widetilde U}}
\newcommand{\Ub} {\mathbb{U}}
\newcommand{\Hc} {\mathcal{H}}
\newcommand{\Oc} {\mathcal{O}}
\newcommand{\Rc} {\mathcal{R}}
\newcommand{\Hbulk} {\mathcal{H}^\text{bulk}}

\begin{document}
\thispagestyle{empty}
\def\thefootnote{\fnsymbol{footnote}}
\begin{flushright}
KCL-MTH-06-10\\
  hep-th/0608184
\end{flushright}
\vskip 1.0em
\begin{center}\LARGE
The logarithmic triplet theory with boundary
\end{center}\vskip 1.5em
\begin{center}\large
  Matthias R. Gaberdiel%
  $^{a}$\footnote{Email: {\tt gaberdiel@itp.phys.ethz.ch}}
  and
  Ingo Runkel%
  $^{b}$\footnote{Email: {\tt ingo.runkel@kcl.ac.uk}}%
\end{center}
\begin{center}\it$^a$
Institute for Theoretical Physics, ETH Z\"urich \\
8093 Z\"urich, Switzerland
\end{center}
\begin{center}\it$^b$
  Department of Mathematics, King's College London \\
  Strand, London WC2R 2LS, United Kingdom  
\end{center}
\vskip .5em
\begin{center}
  August 2006
\end{center}
\vskip 1em
\begin{abstract}
The boundary theory for the $c=-2$ triplet model is investigated in
detail. In particular, we show that there are four different boundary
conditions that preserve the triplet algebra, and check the
consistency of the corresponding boundary operators 
by constructing their OPE coefficients explicitly. We also compute the
correlation functions of two bulk fields in the presence of a
boundary, and verify that they are consistent with factorisation.
\end{abstract}

\setcounter{footnote}{0}
\def\thefootnote{\arabic{footnote}}

\sect{Introduction}

`Rational' logarithmic conformal field theories are logarithmic
conformal field theories that behave in many respects like ordinary
rational conformal field  theories. They are, however, not rational in 
the strict sense since they contain indecomposable representations
(that typically lead to correlation functions with logarithmic branch
cuts). As such these theories provide an interesting class of models
that allow one to probe how far methods developed for standard
rational conformal field theory may in fact be applicable in a wider 
context. In this paper we study the boundary theory for one such
logarithmic theory in detail, the `rational' triplet theory at
$c=-2$. As we shall see, certain aspects of our construction work as
in the usual rational case, but there are also interesting
differences.

The first example of a (non-rational) logarithmic
conformal field theory was already found some time ago in
\cite{Gurarie:1993xq} (see also \cite{RSal92}), and the first
`rational' example (that shall also concern us in this paper) was
constructed in  \cite{GKau96b}; for some recent reviews see 
\cite{Flohr:2001zs,Gaberdiel:2001tr,Kawai:2002fu}. From 
a physics point of view, logarithmic conformal field theories
appear naturally in various models of statistical physics, for example
in  the theory of (multi)critical polymers \cite{Sal92,Flohr95,Kau95},
percolation \cite{Watts96,Flohr:2005ai}, and various critical
(disordered) models 
\cite{CKT95,MSer96,CTT98,Gurarie:1999yx,Ruelle,deGier:2003dg,Piroux:2004vd,%
Moghimi-Araghi:2004wg,Jeng:2006tg}; 
a family of integrable lattice models with logarithmic critical
behaviour has also recently been found in 
\cite{Pearce:2006sz}.
There have been applications in string theory, in particular in
the context of D-brane recoil \cite{KMa95,PT,KMW,Lambert:2003zr}, and
in pp-wave backgrounds \cite{Bakas:2002qh}. 
Logarithmic vertex operator algebras have finally
attracted some attention in mathematics 
\cite{Milas:2001bb,Miyamoto,Huang:2003za}.  
Most work has been done on the $c=-2$ triplet theory, but logarithmic
conformal field theories have also appeared in other contexts, in
particular for theories with super group symmetries,
see for example \cite{RSal92,MSer96,Gurarie:2004ce,Schomerus:2005bf},
as well as in other classes of models, for example  
\cite{Gaberdiel:2001ny,Fjelstad:2002ei,Lesage:2002ch,Rasmussen:2005hj}.

Many structural aspects of logarithmic conformal field theories have
been studied in detail, but there are still a number of issues
that have not yet been satisfactorily understood. One concerns the
structure of the boundary theory that is of some importance since
lattice calculations typically involve boundaries (see for example
\cite{IPRH,Pearce:2006sz,Jeng:2006tg}). Little is known in general
about the structure of a logarithmic boundary theory
\cite{Moghimi-Araghi:2000cx,Kogan:2000fa,Ishimoto:2001jv,Kawai:2001ur}, 
and even for the 
simplest `rational' theory, the triplet theory at $c=-2$, the situation
is somewhat unclear. Various attempts to analyse the boundary
conditions for this theory have been made in the past 
\cite{Kogan:2000fa,Ishimoto:2001jv,Kawai:2001ur,Ruelle,%
Bredthauer:2002ct,Bredthauer:2002xb}, 
but these are partially conflicting and no clear consensus
seems to have emerged. 

In this paper we remedy this situation by studying the boundary theory
of the $c=-2$ triplet model from first principles. We begin by 
constructing the boundary states using the free fermionic formulation
of the theory, and find agreement with the boundary states of
\cite{Kawai:2001ur} and \cite{Ruelle}. These boundary states
satisfy the Cardy condition, and we can read off from this analysis
the boundary field content on the various boundaries and between
different boundary conditions. With this information
one can then analyse
whether these boundary fields define indeed a consistent associative
algebra. This amounts to constructing the relevant boundary OPE
coefficients that have to satisfy the usual crossing relations; 
for the most interesting boundary fields 
(namely those that are primary with respect to the free fermion modes)
we construct these OPE coefficients explicitly. We also determine the
leading bulk-boundary  OPE coefficients; this
allows us to verify that our boundary conditions satisfy some of the
factorisation constraints, {\it i.e.}\  that they are in fact
compatible with the bulk theory of \cite{Gaberdiel:1998ps}.
Many of these checks are highly non-trivial, and taken together they
give very strong support to the assertion that the boundary conditions
we construct are in fact consistent. At least for this logarithmic
triplet theory we have therefore constructed the 
boundary theory in some detail. It would be interesting to understand
which of the features of our construction will generalise for  other 
`rational' logarithmic theories. 
\medskip

The paper is organised as follows. In the next section we give a
brief self-contained summary of our results. The boundary states are
constructed in section~3 where we also determine the open string
spectra. In section~4 we study the associativity of the boundary
operators and construct the relevant OPE coefficients
explicitly. The bulk-boundary OPE coefficients are determined in
section~5 where we also use them to check the factorisation
constraint involving two bulk fields on the upper half
plane. Section~6 contains our conclusions and a brief outlook. Many of
the technical details of our calculations (as well as some of the more
mathematical subtleties) are explained in various appendices.

\sect{Summary of results}\label{summary}

Before explaining the detailed construction of the boundary theory for
the logarithmic $c=-2$ model, let us briefly summarise our main
findings. In order to be self-contained we begin by reviewing the
structure of the bulk theory, following
\cite{Kausch91,Flohr95,Kau95,GKau96b,Gaberdiel:1998ps,Kausch:2000fu}.

\subsection{The chiral structure}

The symmetry algebra (or chiral algebra) of our theory is the
triplet algebra that was defined in \cite{Kausch91}. It is generated
by  the Virasoro modes $L_n$, and the modes of a triplet of weight 3 
fields $W^a_n$. The commutation relations are given in 
appendix~\ref{app:trip}.

\subsubsection{Symplectic fermions}

The algebra has a free field realisation in terms of a pair of
symplectic fermions $\chi^\alpha$ with $\alpha=\pm$
\cite{Gaberdiel:1998ps,Kausch:2000fu}; these are fermionic fields of
conformal weight $h=1$, whose anti-commutation relations are  
\begin{equation}
  \{ \chi^\alpha_m, \chi^\beta_n \} = m \, d^{\alpha\beta} \, 
  \delta_{m,-n}\ . 
\end{equation}
Here  the anti-symmetric tensor $d^{\alpha\beta}$ is normalised to 
$d^{\pm\mp} = \pm 1$. We also introduce the inverse tensor by 
$d_{\mp\pm} = \pm 1$. The triplet generators can be expressed in terms
of these fermions as 
\be
  L_{-2} \Omega =  
  \frac12 \, d_{\alpha\beta} \, \chi^\alpha_{-1} \chi^\beta_{-1} \Omega  
  \quad , \qquad
  W^a_{-3} \Omega =  
  t^a_{\alpha\beta} \,\chi^\alpha_{-2} \chi^\beta_{-1} \Omega \ , 
\ee
where $\Omega$ is the usual vacuum state that is annihilated by all 
$\chi^\alpha_m \Omega=0$ with $m\geq 0$, and the tensors 
$t^a_{\alpha\beta}$ are defined in (\ref{tten}).

The triplet generators are bosonic generators, and the vacuum
representation $\Hc_\Omega$ of the fermionic generators is therefore
not irreducible with respect to the triplet algebra; instead we
have 
\be
{\cal H}_\Omega = {\cal V}_0 \oplus {\cal V}_1 \ .
\ee
Here ${\cal V}_0$ is the irreducible vacuum representation of the
triplet algebra, and ${\cal V}_1$ is the highest weight representation
whose highest weight states $\psi^\pm = \chi^\pm_{-1} \Omega$
form a doublet of states at conformal weight one (that are mapped into
one another under the action of $W^a_0$ --- see \cite{GKau96b} for
more details). In fact, ${\cal  V}_0$ consists of all bosonic states
in  ${\cal H}_\Omega$, while ${\cal  V}_1$ contains all the fermionic 
states. 

\subsubsection{The relevant representations of the triplet algebra}

The vacuum representation ${\cal H}_\Omega$ is a subrepresentation of
the (chiral) highest weight representation generated by $\omega$,
where $\omega$ is characterised by $\chi^\alpha_m \, \omega = 0$ for
$m>0$ (but we do not assume that $\chi^\alpha_0 \omega = 0$).
The space of ground states of this representation is four-dimensional;
it consists of the two bosonic states $\omega$ and   
$\Omega = \chi^-_0 \chi^+_0 \omega$, as well as 
the two fermionic states $\chi^\pm_0 \omega$. The representation
${\cal H}_\omega$ that is generated by the action of 
the free fermions from $\omega$ decomposes into two indecomposable
representations of the triplet algebra  
\be\label{omegadec}
{\cal H}_\omega = {\cal R}_0 \oplus {\cal R}_1 \ ,
\ee
where ${\cal R}_0$ consists of the bosonic, and ${\cal R}_1$ of the  
fermionic states. The cyclic state of ${\cal R}_0$ is the highest
weight state $\omega$; it is annihilated by all positive triplet modes
as well as $W^a_0$, but satisfies
\begin{equation}
L_0 \, \omega = \Omega \ . 
\end{equation}
It therefore defines an indecomposable but reducible 
highest weight representation of the triplet algebra  
\cite{GabKau96a,GKau96b}. The representation ${\cal R}_1$ on the other 
hand is generated by the action of the triplet generators from 
a cyclic state at $h=1$ that is not highest weight; its structure is
described in detail in appendix~\ref{app:trip}. 
\medskip

The other free fermion representation that is of relevance is the 
$\Zb_2$ twisted representation that is generated from a highest
weight state $\mu$ of conformal dimension $h=-1/8$ by the action of
the half-integer moded symplectic fermions. (Since the triplet
generators are bilinear in the symplectic fermions, these are the only
two fermionic representations that lead to untwisted representations
of the triplet algebra.) In terms of the triplet algebra, the
representation  ${\cal H}_\mu$ generated from $\mu$ decomposes into the
two irreducible triplet representations
\be
{\cal H}_\mu = {\cal V}_{-1/8} \oplus {\cal V}_{3/8} \ .
\ee
These are conventional highest weight representations whose highest
weight states have conformal weight $-1/8$ and $3/8$, respectively. 
The highest weight states of ${\cal V}_{3/8}$ are the doublet 
$\chi^\pm_{-1/2} \mu$  that are mapped into one another under 
the action of the $W^a_0$ modes --- for details see \cite{GKau96b}.  

It is also known that the triplet algebra is $C_2$-cofinite  
\cite{GKau96b,Abe,Carqueville:2005nu} and thus is `rational' in that
it has only finitely many highest weight representations. We mention
in passing that the algebra possesses also 
other representations than the ones discussed above
\cite{Fuchs:2003yu,Feigin:2005xs}; however these additional
representations do not appear in the state spaces of the triplet
theory (for the cylinder as well as for the strip), and we need not
discuss them here.

\subsection{The local theory}

We can construct a consistent local theory out of these
representations \cite{Gaberdiel:1998ps}. The space of states can best
be described in terms of the fermionic description of the theory. To
this end we consider the representations of the free fermions  
\be
{\cal H}_{\bomega} \equiv {\cal H}_\omega \otimes \bar{\cal H}_\omega
\ , \qquad
{\cal H}_{\bmu} \equiv {\cal H}_\mu \otimes \bar{\cal H}_\mu \ , 
\ee
where the barred spaces refer to the anti-chiral degrees of
freedom. In both cases we then restrict to the {\it bosonic} degrees
of freedom 
\begin{eqnarray}
{\cal H}_{\bomega}^{\rm bos} & = &
\left( {\cal R}_{0} \otimes \bar{\cal R}_{0} \right) \oplus
\left( {\cal R}_{1} \otimes \bar{\cal R}_{1} \right) 
\label{ombos} \\
{\cal H}_{\bmu}^{\rm bos} & = & 
\left( {\cal V}_{-1/8} \otimes \bar{\cal V}_{-1/8} \right) \oplus
\left( {\cal V}_{3/8} \otimes \bar{\cal V}_{3/8} \right) \ .
\label{irrdec} 
\end{eqnarray}
The total space of the local theory is finally
\be\label{bulk}
{\cal H}^{\rm bulk} = {\cal H}^{\rm bos}_{\bomega} / \Nc 
\oplus {\cal H}^{\rm bos}_{\bmu} \ ,
\ee
where $\Nc$ is the subrepresentation (with respect to the chiral- and
anti-chiral triplet algebra) of ${\cal H}^{\rm bos}_{\bomega}$ that
is spanned by all states of the form
\be
(\chi^\alpha_0 - \bar\chi^\alpha_0) \, \brho  \ , 
\ee
with $\brho$ an arbitrary vector in
the fermionic subspace of 
${\cal H}_{\bomega}$. The quotienting by $\Nc$ is necessary in order 
to obtain a local theory; in the quotient space we then have
\cite{Gaberdiel:1998ps} 
\be
L_0 \, \bomega = \Omega \otimes \bar\omega + \Nc =
\omega\otimes \bar\Omega + \Nc = \bar{L}_0 \, \bomega 
\ee
since 
\be
\Omega \otimes \bar\omega - \omega\otimes \bar\Omega = 
\left[ (\chi^-_0 - \bar\chi^-_0) \, \chi^+_0 
  - (\chi^+_0 - \bar\chi^+_0) \bar\chi^-_0 \right] 
\omega \otimes \bar\omega \in \Nc \ . 
\ee
Here $\bomega$ denotes the equivalence class that contains 
$\omega\otimes \bar\omega$. The other highest weight state of 
${\cal H}^{\rm bos}_{\bomega}/\Nc$ will be denoted by 
\be
\bOmega = \omega \otimes \bar\Omega + \Nc =
\Omega \otimes \bar\omega + \Nc \ . 
\ee
(Note that $\Omega\otimes\Omega \in \Nc$ 
since it is equal to 
$(\chi^-_0 - \bar\chi^-_0) \bar\chi^-_0 \, 
\bar\chi^+_0\, \chi^+_0 \bomega$.)
In terms of the triplet generators the resulting representation space
is then quite complicated \cite{Gaberdiel:1998ps}: given the
decomposition (\ref{omegadec}) it  contains states from  
${\cal R}_0\otimes \bar{\cal R}_0$ and 
${\cal R}_1\otimes \bar{\cal R}_1$, but these get partially identified
upon quotienting by $\Nc$.

\subsubsection{Amplitudes and operator product expansions}

In the following we shall mainly consider the amplitudes of the fields
$\bomega$ and $\bmu$; the amplitudes for the other fields 
can be obtained from these using the fermionic symmetry. We choose the
conventions\footnote{
  This is different from \cite{Gaberdiel:1998ps}:
  if we denote the fields from \cite{Gaberdiel:1998ps} by 
  $\bomega^{\rm GK}$, $\bOmega^{\rm GK}$ and $\bmu^{\rm GK}$, 
  the relation to the fields used here is 
  $\bomega=\bomega^{\rm GK} + 4 \, \log 2 \, \bOmega^{\rm GK}$,
  $\bOmega=\bOmega^{\rm GK}$ and $\bmu = i\bmu^{\rm GK}$. 
  We also choose ${\cal C}_0=-1$.
  With these conventions the 2-point function of $\bomega$ is
  particularly simple, and the bulk-boundary OPE coefficients of
  $\bmu$, as well as the coefficients of the boundary states, are real.}
\be
  \corr{\bomega (z)} = - 1 
  \quad , \qquad
  \corr{\bomega (z_1)\, \bomega (z_2)} =  
  4 \, \log |z_1{-}z_2| \ .  
\ee
The leading terms in the operator product
expansion of the fields $\bomega$ and $\bmu$ are then
\cite{Gaberdiel:1998ps} 
\be\bearll
\bomega (z) \, \bomega(0) \etb =   -4 \log |z| \Big( \bomega (0) 
 + \log |z| \,  \bOmega (0) \Big)+ \cdots \enl
\bmu(z) \, \bomega(0) \etb =  
- 2 \bigl( 2 \log 2 +  \log |z| \bigr) \bmu (0) + \cdots \enl
\bmu(z) \, \bmu(0) \etb =  
|z|^{\frac{1}{2}}\, 
\Big( - \bomega(0) + 2 \big(2 \log 2 - \log|z|\big)\, \bOmega(0) 
+ \cdots \Big) \ .
\eear\labl{eq:bulk-OPE-result}

\subsection{The boundary theory}\label{sec:summary-bndtheory}

As we shall explain in more detail below, the above triplet theory has
four different boundary states that respect the full triplet
symmetry. All four of them can be most easily described in the free
fermion language: there are two `Neumann' boundary states,
$|\!|{\rm N},\pm\rangle\!\rangle$, that satisfy the gluing
conditions 
\be\label{Nglue}
\left(\chi^\alpha_n + \bar\chi^\alpha_{-n} \right) \,
|\!|{\rm N},\pm\rangle\!\rangle = 0 \ ,
\ee
as well as two `Dirichlet' boundary states, 
$|\!|{\rm D},\pm\rangle\!\rangle$, that are characterised by 
\be\label{Dglue}
\left(\chi^\alpha_n - \bar\chi^\alpha_{-n} \right) \,
|\!|{\rm D},\pm\rangle\!\rangle = 0 \ . 
\ee
The relevant open string spectra are (see section~\ref{sec:bnd-states})
\begin{eqnarray}
\langle\!\langle {\rm N},\pm |\!| q^{L_0+\bar{L}_0 - \frac{c}{12}} 
|\!|  {\rm N},\pm \rangle\!\rangle & = & 
\chi_{{\cal R}_0}(\tilde{q}) \nonumber
\\ 
\langle\!\langle {\rm N},\pm |\!| q^{L_0+\bar{L}_0 - \frac{c}{12}} 
|\!|  {\rm N},\mp \rangle\!\rangle & = & 
\chi_{{\cal R}_1}(\tilde{q})
\nonumber  \\
\langle\!\langle {\rm D},\pm |\!| 
q^{L_0+\bar{L}_0 - \frac{c}{12}} 
|\!|  {\rm D},\pm \rangle\!\rangle & = & 
\chi_{{\cal V}_0}(\tilde{q})
\\
\langle\!\langle {\rm D},\pm |\!| q^{L_0+\bar{L}_0 - \frac{c}{12}} 
|\!|  {\rm D},\mp \rangle\!\rangle & = & 
\chi_{{\cal V}_1}(\tilde{q})
\nonumber \\
\langle\!\langle {\rm N},\pm |\!| q^{L_0+\bar{L}_0 - \frac{c}{12}} 
|\!|  {\rm D},\pm \rangle\!\rangle & = & 
\chi_{{\cal V}_{-1/8}}(\tilde{q}) \nonumber \\
\langle\!\langle {\rm N},\pm |\!| q^{L_0+\bar{L}_0 - \frac{c}{12}} 
|\!|  {\rm D},\mp \rangle\!\rangle & = & 
\chi_{{\cal V}_{3/8}}(\tilde{q})\ , \nonumber 
\end{eqnarray}
where $\tilde{q}$ is the open string loop parameter. These spectra
agree with what was found in \cite{Kawai:2001ur} (see also
\cite{Ruelle}). In fact, we may identify our boundary states with the
four irreducible representations as follows
\be\bearll
|\!| {\cal V}_0 \rangle\!\rangle = |\!|  {\rm D},+ \rangle\!\rangle 
\qquad \etb 
|\!| {\cal V}_1 \rangle\!\rangle = |\!|  {\rm D},- \rangle\!\rangle 
\vspace{0.2cm} \enl
|\!| {\cal V}_{-1/8} \rangle\!\rangle = |\!|  {\rm N},+ \rangle\!\rangle 
\qquad \etb
|\!| {\cal V}_{3/8} \rangle\!\rangle = |\!|  {\rm N},- \rangle\!\rangle 
\ . 
\eear\ee
The open string spectra are then simply given by the fusion rules
\begin{equation}
\langle\!\langle {\cal W}_1 |\!| q^{L_0+\bar{L}_0 - \frac{c}{12}} 
|\!|  {\cal W}_2 \rangle\!\rangle  =  
\sum_{{\cal R}} {\cal N}_{{\cal W}_1 \, {\cal W}_2}{}^{\cal R} \, 
\chi_{\cal R}(\tilde{q}) \ , 
\end{equation}
where ${\cal N}$ denotes the fusion rules that were calculated in 
\cite{GKau96b}. Here ${\cal W}_i$ denotes an irreducible
representation, while ${\cal R}$ runs over all indecomposable
representations. This description of the boundary states is
therefore very reminiscent of the usual rational case.  
\medskip

The boundary operators that live on each of the two 
Dirichlet boundary conditions  lie in the irreducible vacuum
representation ${\cal V}_0$ with highest weight state $\Omega$. For
the two Neumann boundary conditions, on the other hand, the boundary
operators lie in the reducible representation ${\cal R}_0$ whose
ground states are $\Omega$ and $\omega$. We choose again the
convention that $L_0 \omega = \Omega$, and that $\Omega$ acts as the
identity field in boundary correlators. Furthermore, we choose our
normalisation so that 
\be
\langle \omega(x) \, \omega(y) \rangle_{\rm uhp}^{{\rm N},\pm} =  
- 2 \, \log(x-y) \,
\langle \omega(0) \rangle_{\rm uhp}^{{\rm N},\pm} \ ,
\ee
where $x$ and $y$ lie on the real axis that describes the boundary of
the upper half plane (uhp). We have determined the leading
order in the boundary operator product expansion 
\be
\omega(x)\omega(0) = - (\log x)^2\, \Omega(0) 
  - 2 \log x \, \omega(0) + \cdots\ ,
\labl{eq:summary-omom-bnd-ope}
and we check in section~\ref{sec:op-alg-bnd} that these 
boundary operators define indeed an associative algebra.
We also confirm there the consistency with the operator product
expansions of boundary changing operators $\mu$ that interpolate 
between $({\rm D},\pm)$ and $({\rm N},\pm)$ boundaries.

Finally we have determined the bulk-boundary operator product
expansions, and we have found that on the upper half plane the
resulting expressions are 
\be
\bearll
  \bmu(iy)\bnd{\Dr,\eta} \etb
  =~\eta \sqrt{\pi} \, (2y)^{\frac14} \Omega(0) + \dots
  \enl
  \bmu(iy)\bnd{\Nr,\eta} \etb
   =~ - \eta \frac{2}{\sqrt{\pi}}\,  (2y)^{\frac14} \big( \omega(0) +
    (\log(2y) - 2\log 2) \Omega(0) \big) + \dots
  \enl
  \bomega(iy)\bnd{\Dr,\eta} \etb
   =~ - 2 \log(2y) \Omega(0) + \dots
  \enl
  \bomega(iy)\bnd{\Nr,\eta} \etb
   =~ 4 \omega(0) + 2\log(2y) \Omega(0) + \dots \ , 
\eear
\labl{eq:bulk-bnd-OPE-result}
where $\eta=\pm$.
We have checked that with
these conventions, the boundary conditions satisfy the factorisation
constraints that come from considering the amplitude with two bulk
fields on the upper half plane (see section~\ref{sec:bbO+fac}). 

Given that our boundary conditions satisfy all of these consistency
conditions, it is very plausible to believe that they are indeed
consistent boundary conditions of the triplet bulk theory. Thus, at
least for this logarithmic theory, we have managed to construct the
boundary theory. 
\smallskip

The rest of the paper is somewhat more technical; in the following
sections we shall explain in detail how to derive these results and
check their consistency. The first step is the construction of the
boundary states.

\sect{The construction of the boundary states}\label{sec:bnd-states}

\subsection{Ishibashi states}

In the following we want to construct all the boundary states of
the triplet theory that preserve the triplet symmetry (with trivial
gluing automorphism). These boundary states have to lie
in the subspace $B$ of (a suitable completion of) $\Hc^\text{bulk}$
given by
\be
 B = \Big\{ |v\rangle\!\rangle \in \Hc^\text{bulk} \,\Big|\,
(L_m {-} \ol L_{-m})
|v\rangle\!\rangle = 0 = (W_m^a {+} \ol W{}_{-m}^a) |v\rangle\!\rangle
\,;~ m \in \Zb \,,\, a \in \{+,0,-\} \, \Big\} \ .
\labl{eq:bnd-state-space}
For usual rational conformal field theories whose space of states is a
direct sum of tensor products $\Hc_i \otimes \ol \Hc_j$ where $\Hc_i$
and  
$\ol \Hc_j$ are irreducible representations, the space of solutions is
spanned by the Ishibashi states,
of which there is (up to normalisation) one in each sector  
$\Hc_i \otimes \ol \Hc_i$. 
For the triplet theory two new features
appear: some of the representations are indecomposable but reducible,
and the space of states is actually a quotient. 
A sector by sector analysis of possible solutions to the conditions
in \erf{eq:bnd-state-space} (without imposing
any constraints that come from the quotient space ${\cal N}$)
has been carried out in \cite{Bredthauer:2002ct}; the results have
also been compared to what one obtains from requiring the fermionic
symmetries to be preserved \cite{Bredthauer:2002xb}.
\medskip

In the following we shall not try to construct the space
$B \subset \Hc^\text{bulk}$ directly, but first 
consider the Ishibashi states that preserve the full 
fermionic symmetries. 
The relevant gluing conditions for the fermions
that guarantee that the triplet symmetry is preserved are either
`Dirichlet'  
\be
\left( \chi^\pm_n - \bar\chi^\pm_{-n} \right) 
|\!|{\rm D} \rangle\!\rangle = 0  
\labl{eq:D-glue}
or `Neumann'
\be
\left( \chi^\pm_n + \bar\chi^\pm_{-n} \right) 
|\!|{\rm N}\rangle\!\rangle =0 \ , 
\labl{eq:N-glue}
where it is understood that $n\in\Zb$ for the sector
$\Hc_{\bomega}^\text{bos} / \Nc \subset \Hc^\text{bulk}$, 
while $n\in \Zb + \tfrac12$ for 
$\Hc_{\bmu}^\text{bos}$.

Given a state $\brho \in \Hc^{\text{bulk}}$ that is
annihilated by all positive fermion modes, the coherent states
\be\bearll
|\brho\rangle\!\rangle^{{\rm (D)}} \etb = ~
\exp\!\Big(  \sum_{n>0} \frac{1}{n} ( \chi^-_{-n} \bar\chi^+_{-n} - 
\chi^+_{-n} \bar\chi^-_{-n} )  \Big) \brho \enl
|\brho\rangle\!\rangle^{{\rm (N)}} \etb = ~
\exp\!\Big( - \sum_{n>0} \frac{1}{n} ( \chi^-_{-n} \bar\chi^+_{-n} - 
\chi^+_{-n} \bar\chi^-_{-n} )  \Big) \brho 
\eear\labl{eq:coh-states}
satisfy the conditions \erf{eq:D-glue} or \erf{eq:N-glue},
respectively, for $n \neq 0$. Since there are no fermionic zero modes
in $\Hc_{\bmu}^\text{bos}$, we therefore have two solutions to 
\erf{eq:bnd-state-space} by taking $\brho = \bmu$.

The analysis is more complicated if 
$\brho \in \Hc_{\bomega}^\text{bos} / \Nc$, 
since there are zero modes, and we need to impose (\ref{eq:D-glue})
and (\ref{eq:N-glue}) also for these zero modes.\footnote{This seems
to have been overlooked in \cite{Kawai:2001ur}.}
The zero mode constraint
of (\ref{eq:D-glue}) is trivial since the image under 
$\chi^\pm_0 - \bar\chi^\pm_0$ lies in $\Nc$, and is thus automatically
zero in the quotient space. On the other hand, it is clear that
$\bomega$ does {\it not} satisfy the zero mode constraint of 
(\ref{eq:N-glue}) since 
$(\chi^\pm_0 + \bar\chi^\pm_0) \, \bomega = 2 \chi^\pm_0 \bomega \ne 0$. 
(There is no such condition for the Ishibashi states based on
$\bOmega$, since $\bOmega$ is annihilated by all fermionic zero
modes.) Thus \erf{eq:coh-states} gives only three fermionic Ishibashi
states  in the sector 
$\Hc_{\bomega}^\text{bos} / \Nc \subset {\cal H}^{\rm bulk}$.  
So far, we therefore have the five Ishibashi states  
\be\bearll
{\rm (D)}: \qquad \etb |\bmu \rangle\!\rangle^{{\rm (D)}}  \quad
|\bomega\rangle\!\rangle^{{\rm (D)}}  \quad
|\bOmega \rangle\!\rangle^{{\rm (D)}} \enl
{\rm (N)}: \qquad \etb |\bmu \rangle\!\rangle^{{\rm (N)}}
\quad |\bOmega \rangle\!\rangle^{{\rm (N)}}\ . 
\eear\labl{eq:ferm-Ishi}
The space $B$ thus has at least dimension five. We now proceed to
prove that it has exactly dimension five, {\it i.e.}\ that the
Ishibashi states \erf{eq:ferm-Ishi} span $B$. 
\medskip

Every element $|v\rangle\!\rangle$ of $B$ can be written as a direct
sum 
$|v\rangle\!\rangle 
= |v_\omega\rangle\!\rangle\oplus |v_\mu\rangle\!\rangle$, where 
$|v_\om\rangle\!\rangle \in \Hc_{\bomega}^\text{bos} / \Nc$ 
and $|v_\mu\rangle\!\rangle \in \Hc_{\bmu}^\text{bos}$. Since the
action of $L_m$ and $W^a_m$ does not mix these two sectors, it follows
that also $|v_\omega\rangle\!\rangle$ and $|v_\mu\rangle\!\rangle$
must be individually in $B$. It is therefore enough to look for
solutions to \erf{eq:bnd-state-space} in these two sectors of
$\Hc^\text{bulk}$ separately. Accordingly we will 
write $B = B_\omega \oplus B_\mu$. 

Let us start with $\Hc_{\bmu}^\text{bos}$. Recall that this sector
decomposes as 
$( {\cal V}_{-1/8} \otimes \bar{\cal V}_{-1/8} ) \oplus
( {\cal V}_{3/8} \otimes \bar{\cal V}_{3/8} )$ into triplet 
representations. These are all irreducible, and so this sector
gives rise to exactly two Ishibashi states. We have already found
two states in $B_\mu$, namely 
$|\bmu \rangle\!\rangle^{{\rm (D)}}$ and 
$|\bmu \rangle\!\rangle^{{\rm (N)}}$, and hence these already
span $B_\mu$ (each of these two states is a 
linear combinations of the Ishibashi states corresponding to the two 
triplet representations).

The analysis in the sector $\Hc_{\bomega}^\text{bos} / \Nc$ is more
involved. We already know that the three states
$|\bomega\rangle\!\rangle^{{\rm (D)}}$,
$|\bOmega \rangle\!\rangle^{{\rm (D)}}$ and
$|\bOmega \rangle\!\rangle^{{\rm (N)}}$ lie in $B_\omega$. Let
now $|b\rangle\!\rangle$ be an arbitrary element in $B_\om$.
We can expand $|b\rangle\!\rangle$ as
\be
  |b\rangle\!\rangle  = \sum_{k=0}^\infty v^{(k)} \ ;
  \qquad v^{(k)} ~\text{has~grade}~(k,k) \ .
\ee
In the quotient space $\Hc_{\bomega}^\text{bos} / \Nc$, 
we can replace any zero mode $\bar\chi_0^a$ by $\chi_0^a$. Then it is
easy to see that the two lowest grades have to be of the
form
\be
  v^{(0)} = b^\omega \bomega + b^\Omega \bOmega \ ,
    \qquad
  v^{(1)} = b^\omega_{\alpha\beta} \chi^\alpha_{-1} 
  \bar\chi^\beta_{-1} \bomega + b^\Omega_{\alpha\beta} \chi^\alpha_{-1} 
  \bar\chi^\beta_{-1} \bOmega  
\ee
for some constants $b^\omega$, $b^\Omega$, $b^\omega_{\alpha\beta}$,
and $b^\Omega_{\alpha\beta}$. Consider the following linear combination
of Ishibashi states,
\be
  |b'\rangle\!\rangle  = |b\rangle\!\rangle
  - b^\omega |\bomega\rangle\!\rangle^{{\rm (D)}}
  - \tfrac12 b^\Omega \big(
|\bOmega \rangle\!\rangle^{{\rm (D)}} +
|\bOmega \rangle\!\rangle^{{\rm (N)}} \big)
  - \tfrac12 b^\Omega_{-+} \big(
|\bOmega \rangle\!\rangle^{{\rm (D)}} -
|\bOmega \rangle\!\rangle^{{\rm (N)}} \big) \ .
\labl{eq:bprime-def}
We will prove below that $|b'\rangle\!\rangle = 0$ and hence that
every element in $B_\omega$ is a linear combination of 
the three fermionic states already found. This then implies
$\dim B_\omega = 3$, as claimed.
\medskip

Let us decompose $|b'\rangle\!\rangle  = \sum_{k=0}^\infty u^{(k)}$, 
where again $u^{(k)}$ has grade $(k,k)$.
The linear combination \erf{eq:bprime-def} is chosen such that
$u^{(0)} = 0$. 
We proceed by an induction argument. Suppose we know that
$u^{(k)}=0$ for $k \le N{-}1$ for some $N>0$. Then 
the conditions in \erf{eq:bnd-state-space} imply, for $m>0$,
\be
  L_m u^{(N)} = \ol L_{-m} u^{(N-2m)} = 0 \ ,  \qquad 
  \ol L_m u^{(N)} = L_{-m} u^{(N-2m)} = 0 \ ,
\ee
where in each chain of equalities, the first equality
follows since $(L_m{-}\ol L_{-m})|b'\rangle\!\rangle =0$, while 
the second follows from $u^{(k)}=0$ for $k < N$.
Similarly one sees that $W^a_m u^{(N)} = 0 = \ol W{}^a_m u^{(N)}$. 
Thus $u^{(N)}$ must be a triplet primary. The only triplet primary
states in $\Hc_{\bomega}^\text{bos} / \Nc$ are at grades\footnote{ 
This is clear for the representation 
$\Hc_{\bomega}^\text{bos} = 
(\Rc_0 \otimes \Rc_0) \oplus (\Rc_1 \otimes \Rc_1)$, but some
care has to be taken for the quotient $\Hc_{\bomega}^\text{bos} / \Nc$.
Note however, that $\Hc_{\bomega}^\text{bos} / \Nc$ is in particular a
representation of the holomorphic copy of the triplet algebra in the
bulk. All its representations (with integer generalised $L_0$-weights)
are known, and they have highest weight states only for grades 0 or 1.
The same applies to the anti-holomorphic copy of the triplet algebra.}
$(0,0)$, $(1,0)$, $(0,1)$ and $(1,1)$. Thus we only need to consider
the case $N=1$, for which $u^{(1)}$ is given by
\be
  u^{(1)} = f^\omega_{\alpha\beta} \chi^\alpha_{-1} 
  \bar\chi^\beta_{-1} \bomega + f^\Omega_{\alpha\beta} \chi^\alpha_{-1} 
  \bar\chi^\beta_{-1} \bOmega  \ ; 
  \qquad 
   f^\omega_{\alpha\beta} = b^\omega_{\alpha\beta}
     - b^\omega d_{\alpha\beta} \ , \quad 
   f^\Omega_{\alpha\beta} = b^\Omega_{\alpha\beta}
     - b^\Omega_{-+} d_{\alpha\beta} \ .
\ee
By construction we have $f^\Omega_{-+}=0$. Since the action of
$L_1$, $W^a_1$, 
$\ol L_1$, $\ol W{}^a_1$ on a grade $(1,1)$ state contributes
one zero mode, we see that $\chi^\alpha_{-1} 
\bar\chi^\beta_{-1} \bOmega$ is triplet primary. Similarly, one can also
check that $\chi^\alpha_{-1} \bar\chi^\beta_{-1} \bomega$ is {\em not}
triplet primary. It therefore follows that
$f^\omega_{\alpha\beta}=0$. Finally, we impose the zero mode condition 
$(W^a_0 + \ol W{}^a_0) u^{(1)}=0$. 
In terms of fermion modes, this is equivalent to demanding
\be
  t^a_{\rho\sigma}\big(
  \chi^\rho_{-1} \chi^\sigma_1 +
  \bar\chi^\rho_{-1} \bar\chi^\sigma_1 \big)
  f^\Omega_{\alpha\beta} \chi^\alpha_{-1} 
  \bar\chi^\beta_{-1} \bOmega  = 0  \ .
\labl{eq:W0+W0cond}
Expressing this in terms of negative fermion modes only, one finds
that \erf{eq:W0+W0cond} is equivalent to
$t^a_{\gamma \sigma} d^{\sigma\alpha} f^\Omega_{\alpha\delta}
  + t^a_{\delta \sigma} d^{\sigma\beta} f^\Omega_{\gamma\beta} = 0$
for all $a = 0,\pm$ and $\gamma,\delta \in \{ \pm \}$. This in turn
can be reduced to $f^\Omega_{++} = f^\Omega_{--} = 0$ and
$f^\Omega_{+-} + f^\Omega_{-+} = 0$. As we have already found that 
$f^\Omega_{-+}=0$ we can conclude that 
$f^\Omega_{\alpha\beta}=0$. Thus also $u^{(1)}=0$.
Altogether we therefore find that indeed $|b'\rangle\!\rangle =0$.
\medskip

This completes the proof that the five Ishibashi states
\erf{eq:ferm-Ishi} are a basis of $B$.
Naively, we would therefore expect that there are five boundary states
for the triplet theory. However, as we shall see, there are only
four linear combinations that actually define boundary conditions that
are consistent with the Cardy constraint.

\subsection{Calculation of cylinder diagrams}

In order to determine the consistent boundary states we can use the
Cardy condition. This requires calculating the cylinder amplitude
between the different Ishibashi states given above. For the
logarithmic theory we are considering here, there is a subtlety 
regarding the definition of the inner product between states, and we
therefore need to explain carefully what we need to calculate.

As is explained in appendix~\ref{app:local-coord}, every conformal
field theory (logarithmic or not) possesses a bilinear form on the
bulk space; the cylinder amplitude is then simply the bilinear form
evaluated on the two boundary states, {\it i.e.}
\be
\langle\!\langle {\rm B}_1 |\!| 
q^{\frac{1}{2}(L_0+\bar{L}_0) - \frac{c}{24}} 
|\!|  {\rm B}_2 \rangle\!\rangle = 
B\left( |\!|  {\rm B}_1 \rangle\!\rangle, 
q^{\frac{1}{2}(L_0+\bar{L}_0) - \frac{c}{24}} 
|\!|  {\rm B}_2 \rangle\!\rangle \right)\ . 
\ee
With this prescription and the definitions of appendix~\ref{app:local-coord} 
(recall that $B(\bomega,\bOmega)=-1$ and $B(\bmu,\bmu)=1$), it is then  
straightforward to calculate the overlaps between the Ishibashi states
\be\begin{array}{lll}\displaystyle
{} {}^{\rm (D)} \langle\!\langle \bmu | 
q^{\frac{1}{2}(L_0 + \bar{L}_0) - \frac{c}{24}} 
|\bmu \rangle\!\rangle^{{\rm (D)}} \etb=~ 
{}^{\rm (N)} \langle\!\langle \bmu | 
q^{\frac{1}{2}(L_0 + \bar{L}_0) - \frac{c}{24}} 
|\bmu \rangle\!\rangle^{{\rm (N)}} \etb=~ f_4(q)^2 =~
f_2(\tilde{q})^2 
\enl
{} {}^{\rm (D)} \langle\!\langle \bmu | 
q^{\frac{1}{2}(L_0 + \bar{L}_0) - \frac{c}{24}} 
|\bmu \rangle\!\rangle^{{\rm (N)}} \etb=~ 
{}^{\rm (N)} \langle\!\langle \bmu | 
q^{\frac{1}{2}(L_0 + \bar{L}_0) - \frac{c}{24}} 
|\bmu \rangle\!\rangle^{{\rm (D)}} \etb=~ f_3(q)^2  =~
f_3(\tilde{q})^2 
\enl
{}^{\rm (D)} \langle\!\langle \bomega | 
q^{\frac{1}{2}(L_0 + \bar{L}_0) - \frac{c}{24}} 
|\bOmega \rangle\!\rangle^{{\rm (D)}} \etb=~ 
{}^{\rm (D)} \langle\!\langle \bOmega | 
q^{\frac{1}{2}(L_0 + \bar{L}_0) - \frac{c}{24}} 
|\bomega \rangle\!\rangle^{{\rm (D)}} \etb=~ - f_1(q)^2 
 =~   i \tilde{\tau} f_1(\tilde{q})^2 
\enl
{} {}^{\rm (D)} \langle\!\langle \bomega | 
q^{\frac{1}{2}(L_0 + \bar{L}_0) - \frac{c}{24}} 
|\bOmega \rangle\!\rangle^{{\rm (N)}} \etb=~ 
{}^{\rm (N)} \langle\!\langle \bOmega | 
q^{\frac{1}{2}(L_0 + \bar{L}_0) - \frac{c}{24}} 
|\bomega \rangle\!\rangle^{{\rm (D)}} \etb=~  - \frac{1}{2} f_2(q)^2
=~  - \frac{1}{2} f_4(\tilde{q})^2
\enl
{} {}^{\rm (D)} \langle\!\langle \bomega | 
q^{\frac{1}{2}(L_0 + \bar{L}_0) - \frac{c}{24}} 
|\bomega \rangle\!\rangle^{{\rm (D)}} \etb=~ 
- 2\pi i \, \tau\, f_1(q)^2 \makebox[0pt][l]{
$~=~  2\pi \, f_1(\tilde{q})^2\ ,$}  
\eear\labl{eq:ish-overlap}
while all other overlaps vanish. 
Here $\tilde{q}=e^{2\pi i \tilde{\tau}}$ is the open string loop
parameter, with $\tilde{\tau}=-1/\tau$. The relevant theta functions
$f_i$, as well as their behaviour under modular transformations, are
given in appendix~\ref{app:theta}.

\subsection{Solution of Cardy condition}\label{sec:cardy-sol}

Using the simple modular transformation properties stated in
\erf{eq:ish-overlap}, it is easy to check that the following four boundary
states satisfy the Cardy condition, {\it i.e.}\ give overlaps that
can be interpreted as open string state spaces
({\it cf.}\ \cite{Kawai:2001ur}):
\bea
|\!| {\rm D}, \pm \rangle\!\rangle  =  
\frac{1}{{2}} |\bmu \rangle\!\rangle^{{\rm (D)}} \mp 
\frac{1}{\sqrt{4\pi }} \,
|\bomega \rangle\!\rangle^{{\rm (D)}} 
\enl
|\!| {\rm N}, \pm \rangle\!\rangle  =  
 |\bmu \rangle\!\rangle^{{\rm (N)}} \pm \sqrt{4\pi } \,
|\bOmega \rangle\!\rangle^{{\rm (N)}} \ .
\eear\labl{eq:bnd-states}
The overlaps of these branes are then
\be\bearll
\langle\!\langle {\rm D},\pm |\!| 
q^{\frac{1}{2}(L_0+\bar{L}_0) - \frac{c}{24}} 
|\!|  {\rm D},\pm \rangle\!\rangle  \etb=~  
\frac{1}{4} f_2(\tilde{q})^2 + \frac{1}{2} f_1(\tilde{q})^2 
= \chi_{{\cal V}_0}(\tilde{q})
\enl
\langle\!\langle {\rm D},\pm |\!| 
q^{\frac{1}{2}(L_0+\bar{L}_0) - \frac{c}{24}} 
|\!|  {\rm D},\mp \rangle\!\rangle  \etb=~  
\frac{1}{4} f_2(\tilde{q})^2 - \frac{1}{2} f_1(\tilde{q})^2 
= \chi_{{\cal V}_1}(\tilde{q})
\enl
\langle\!\langle {\rm N},\pm |\!| 
q^{\frac{1}{2}(L_0+\bar{L}_0) - \frac{c}{24}} 
|\!|  {\rm N},\pm \rangle\!\rangle  \etb=~  
f_2(\tilde{q})^2  = \chi_{{\cal R}_0}(\tilde{q}) 
\enl
\langle\!\langle {\rm N},\pm |\!| 
q^{\frac{1}{2}(L_0+\bar{L}_0) - \frac{c}{24}} 
|\!|  {\rm N},\mp \rangle\!\rangle  \etb=~  
f_2(\tilde{q})^2  = \chi_{{\cal R}_1}(\tilde{q})
\enl
\langle\!\langle {\rm D},\pm |\!| 
q^{\frac{1}{2}(L_0+\bar{L}_0) - \frac{c}{24}} 
|\!|  {\rm N},\pm \rangle\!\rangle \etb=~ 
\frac{1}{2} f_3(\tilde{q})^2 + \frac{1}{2} f_4(\tilde{q})^2 
= \chi_{{\cal V}_{-1/8}}(\tilde{q}) \vspace*{0.1cm} \\
\langle\!\langle {\rm D},\pm |\!| 
q^{\frac{1}{2}(L_0+\bar{L}_0) - \frac{c}{24}} 
|\!|  {\rm N},\mp \rangle\!\rangle \etb=~  
\frac{1}{2} f_3(\tilde{q})^2 - \frac{1}{2} f_4(\tilde{q})^2 
= \chi_{{\cal V}_{3/8}}(\tilde{q})\ ,
\eear\labl{eq:bnd-overlap}
where $\tilde{q}$ is the open string loop parameter.
This is in agreement with the results of \cite{Kawai:2001ur,Ruelle} and
reproduces our claim of section~\ref{sec:summary-bndtheory}. 
The character identities for the triplet representations were first
derived in \cite{Flohr95,Kau95}.

Note that the identification of characters with representations is 
not unique; indeed one has to take into account the ambiguity
\be
  \chi_{\Rc_0}(q) = \chi_{\Rc_1}(q) = 
  2\chi_{\Vc_0}(q) + 2\chi_{\Vc_1}(q) \ .
\ee
Comparing with the overlaps \erf{eq:bnd-overlap} we see that the
representations formed by the various open state spaces are uniquely
determined by the characters in all cases except for the four
overlaps involving only $(\Nr,\pm)$-boundary conditions. In these
cases the assignment of representations to the open state spaces
as suggested by the notation used in \erf{eq:bnd-overlap} is an
{\em ansatz}, which will be subjected to strong consistency checks in
the following. 
We have also used that the overlap between $(\Nr,\eta)$ and 
$(\Nr,\eta)$ has opposite fermion number relative to that between 
$(\Nr,\eta)$ and $(\Nr,-\eta)$; if one is to be identified with 
${\cal R}_0$ the other must then be ${\cal R}_1$. 
\medskip

There are at least two ways to arrive at the boundary states
\erf{eq:bnd-states}. The first starts from the formulation of
the boundary conditions in terms of symplectic fermions as in
\erf{eq:D-glue} and \erf{eq:N-glue}. 
One then demands the boundary states (rather than only the
specific basis of Ishibashi states \erf{eq:ferm-Ishi}) to be compatible
with these fermionic gluing conditions, which amounts to taking a
boundary state to be either a linear combination of only
the $(\Dr)$ Ishibashi states in \erf{eq:ferm-Ishi}, or of only the 
$(\Nr)$ Ishibashi states. Imposing the Cardy condition on this
restricted ansatz, and demanding that the resulting set of boundary
states cannot be written as a non-trivial non-negative integer
combination of another set, one arrives at \erf{eq:bnd-states}.

Second, as was done in \cite{Kawai:2001ur}, one can demand the
existence of a boundary state $|\!|\Vc_0\rangle\!\rangle$, 
for which the space of open string states form the vacuum 
representation $\Vc_0$ of the triplet algebra. Finding a maximal
set of fundamental boundary states compatible with 
$|\!|\Vc_0\rangle\!\rangle$ then also leads to \erf{eq:bnd-states}. 

However, this is not to say that \erf{eq:bnd-states} is the only
set of boundary states consistent with the Cardy condition. 
For example, one can check that
$|\bmu \rangle\!\rangle^{{\rm (D)}} 
+ \alpha_k |\bOmega \rangle\!\rangle^{{\rm (D)}} 
+ \beta_k |\bOmega \rangle\!\rangle^{{\rm (N)}}$
and
$|\bmu \rangle\!\rangle^{{\rm (N)}} 
+ \gamma_k |\bOmega \rangle\!\rangle^{{\rm (D)}} 
+ \delta_k |\bOmega \rangle\!\rangle^{{\rm (N)}}$
with $k=1,2$ and $\alpha_k,\beta_k,\gamma_k,\delta_k$ any choice
of complex coefficients (such that the resulting four vectors
are still linearly independent) also gives four boundary states
solving the Cardy condition. 
Of course, the Cardy condition is only one of many necessary conditions
for a consistent boundary theory, and one should therefore not expect
that it alone leads to a unique set of boundary states. 
In the following we 
shall only consider the ansatz \erf{eq:bnd-states}, and show
that it passes a large number of additional consistency checks.

\medskip

In any case, given any four linearly independent and consistent
boundary states, it is clear that one cannot construct a fifth one.
For, if there were five linearly independent
boundary states, there would be two boundary states that involve 
$|\bOmega\rangle\!\rangle^{\rm (D)}$ and 
$|\bomega\rangle\!\rangle^{\rm (D)}$, respectively. Their relative
overlap then leads to a term proportional 
to $f_1(q)^2$ in the closed string channel, which gives rise to 
$\tilde\tau\, f_1(\tilde{q})^2$ in the open string; this does not have  
an interpretation as an open string trace. The fact that there are
four boundary states also ties in nicely with the fact that only four
of the five chiral torus amplitudes of \cite{FG} correspond to 
characters of highest weight representations. 
Some aspects of the modular properties of the triplet theory were
also studied in \cite{Flohr:1996vc,Feigin:2005zx,Fuchs:2006nx}.

\subsection{One point functions}

It is straightforward to deduce from these boundary states the
expressions for the bulk one point functions. In fact, the 
above boundary states are defined on the disc, and thus we can read
off the disc one-point functions from them directly
\be
\bearll
  \langle \bOmega(0) \rangle_{\rm disc}^{{\rm D},\eta} 
          = \frac{\eta}{\sqrt{4\pi}}
  \qquad\qquad\etb  
  \langle \bOmega(0) \rangle_{\rm disc}^{{\rm N},\eta} = 0
  \enl
  \langle \bomega(0) \rangle_{\rm disc}^{{\rm D},\eta} = 0
  \etb  
  \langle \bomega(0) \rangle_{\rm disc}^{{\rm N},\eta} 
          = - \eta\, \sqrt{4\pi}
  \enl
  \langle \bmu(0) \rangle_{\rm disc}^{{\rm D},\eta} 
          = \frac12
  \etb  
  \langle \bmu(0) \rangle_{\rm disc}^{{\rm N},\eta} = 1 \ . 
\eear\ee

Given the one point functions on the disc, we can apply a 
conformal transformation to obtain the one-point functions on the 
upper half plane (uhp). For the $\bOmega$ and $\bmu$ field this is 
straightforward, but for $\bomega$ we have to bear in mind that 
it is not an eigenvector of $L_0$ (or $\bar{L}_0$). In particular, we
therefore have for any boundary condition $b$
\be\bearll
  \langle \bOmega(iy) \rangle_{\rm uhp}^{b} 
  \etb= \lambda \cdot \langle \bOmega(0) \rangle_{\rm disc}^{b}  
  \enl
  \langle \bomega(iy) \rangle_{\rm uhp}^{b} \etb=
  \lambda \cdot \big( \langle \bomega(0) \rangle_{\rm disc}^{b} 
    - 2 \log(2y) \langle \bOmega(0) \rangle_{\rm disc}^{b} \big)
  \enl
  \langle \bmu(iy) \rangle_{\rm uhp}^{b} \etb=
  \lambda \cdot (2y)^{\frac14} \langle \bmu(0) \rangle_{\rm disc}^{b}
  \ ,
\eear\ee
where $\lambda$ is a constant that describes the relative
normalisation of the two amplitudes. We therefore obtain 
\be
\bearll
  \langle \bOmega(iy) \rangle_{\rm uhp}^{{\rm D},\eta}  
  = \lambda\cdot \frac{\eta}{\sqrt{4\pi}} 
  \qquad\qquad\qquad \etb  
  \langle \bOmega(iy) \rangle_{\rm uhp}^{{\rm N},\eta} = 0 
  \enl
  \langle \bomega(iy) \rangle_{\rm uhp}^{{\rm D},\eta} = 
   - 2 \, \lambda \cdot \log(2 y)\, \frac{\eta}{\sqrt{4\pi}}  
  \etb  
  \langle \bomega(iy) \rangle_{\rm uhp}^{{\rm N},\eta} = 
         - \lambda\cdot \eta\sqrt{4\pi}
  \enl
  \langle \bmu(iy) \rangle_{\rm uhp}^{{\rm D},\eta} = 
      \lambda\cdot  \frac12 \, (2y)^{\frac{1}{4}}
  \etb  
  \langle \bmu(iy) \rangle_{\rm uhp}^{{\rm N},\eta} = 
      \lambda\cdot  (2y)^{\frac{1}{4}} \ .
\eear\labl{eq:bulk-1pt-uhp}

\sect{Operator algebra on the boundary}\label{sec:op-alg-bnd}

The above boundary states determine the triplet representations that
appear in the various open string spectra, 
up to the ambiguity 
$\chi_{\Rc_0} = \chi_{\Rc_1} = 2\chi_{\Vc_0} + 2 \chi_{\Vc_1}$.
As already mentioned in section \ref{sec:cardy-sol},
we make the ansatz that the space of states on a strip with
$(\Nr,\pm)$-boundary conditions form the representations
$\Rc_0$ and $\Rc_1$ as suggested by the notation used in
\erf{eq:bnd-overlap}. 

Now we want to check
that these boundary operators form indeed an associative
algebra. For the case of the $(\Dr,\pm)$-boundary conditions, the open
string spectrum is just the vacuum representation of the triplet
algebra which is associative (since the triplet algebra defines a
consistent vertex operator algebra). The situation for the 
$(\Nr,\pm)$-boundary is however more interesting.

\subsection{The boundary OPE $\om(x)\om(0)$}\label{sec:bndOPE-oo}

{}From the analysis of the cylinder partition functions 
in section \ref{sec:cardy-sol} we derived the ansatz 
that the space of boundary
fields on the $(\Nr,\pm)$-boundary forms the representation
$\Rc_0$ under the action of the triplet algebra. At level $0$
we thus have two linearly independent states, $\omega$
and $\Omega = L_0 \omega$. The field $\Omega(x)$ will be
the identity field on the $(\Nr,\pm)$-boundary, and according
to the discussion in appendix \ref{app:2pt-uhp} we choose
$\omega$ such that it has the following 2-point function on
the upper half plane
\be
  \langle \omega(x) \omega(y) \rangle_\text{uhp}^{\Nr,\eta} = -2
                \log(x{-}y) 
  \langle \omega(0) \rangle_\text{uhp}^{\Nr,\eta} 
  \ , \quad \text{where} \quad \eta = \pm \ , \qquad x>y \ .
\labl{eq:om-bnd-2pt}
The boundary 1-point function 
$\langle \omega(0) \rangle_\text{uhp}^{\Nr,\eta}$ 
(not to be confused with the bulk 1-point function
$\langle \bomega(iy) \rangle_\text{uhp}^{\Nr,\eta}$ in 
\erf{eq:bulk-1pt-uhp}) is independent of the insertion point, and its 
value will be determined in section \ref{sec:b1-uhp} below.  

In order to characterise higher order corrections we need to introduce
some notation. In the following $\orl(x^\ell)$ will denote the functions
$f(x)$ for which there is an $n>0$ such that 
$f(x) / ( x^\ell \cdot \log(x)^n ) 
\overset{x \rightarrow 0}{\longrightarrow} 
{\rm const}$. Furthermore, by $\orl[h,x^\ell]$ we refer to a state
that  can be written in the form $\sum_i f_i(x) v_i$ such that each  
$f_i(x)$ is of order $\orl(x^\ell)$, and each vector $v_i$ has
(generalised) $L_0$-weight greater or equal to $h$.

With these conventions we now make the ansatz for the 
leading order of the boundary OPE of $\omega$  
\be
  \omega(x)\omega(0) = f(x) \Omega(0) + g(x) \omega(0) + \orl[h{=}1,x] \ .
\labl{eq:om-bOPE-ansatz}
The functions $f(x)$ and $g(x)$ can be fixed up to constants by
requiring compatibility with the action of $L_0$ as follows. Acting
with $L_0$ on both sides of \erf{eq:om-bOPE-ansatz} gives 
\be
  \big( \Omega(x) + x \tfrac{\partial}{\partial x} \omega(x) \big) 
       \omega(0) + \omega(x) \Omega(0) = g(x) \Omega(0) + \orl[h{=}1,x] \ .
\ee
Substituting the ansatz for the OPE and using that $\Omega(x)$ is the
identity field results in 
\be
  ( x g'(x) + 2 ) \omega(0) + x f'(x) \Omega(0) + \orl[h{=}1,x] 
         = g(x) \Om(0) + \orl[h{=}1,x] \ .
\ee
Comparing coefficients gives two first order differential equations
for $f$ and $g$ which are solved by $g(x) = - 2 \log x + C_1$ and 
$f(x) = - (\log x)^2 + C_1 \log x + C_2$ for some constants $C_1$, $C_2$.
Applying the OPE to the 2-point function in \erf{eq:om-bnd-2pt} further 
determines $C_1 = 0$ so that altogether 
\be
  \omega(x)\omega(0) = \big(C_\om^{\eta} - (\log x)^2\big) \Omega(0) 
  - 2 \log x \, \omega(0) + \orl[h{=}1,x] \ .
\labl{eq:omom-bnd-ope-aux1}
Here $C_\om^{\eta}$ is a constant that potentially depends on the
boundary condition 
$(\Nr,\eta)$. To determine this constant one could try
to carry out a crossing calculation for the correlator of four
$\omega$-fields on the upper half plane. This is done in appendix 
\ref{app:alg-assoc} and turns out not to pose any restrictions on
$C_\om^{\eta}$. We will fix this constant in the next section when
looking at boundary changing fields.

\subsection{The boundary OPE $\mu(x)\om(0)$} 

Up to now we have only considered boundary preserving 
operators. According to \erf{eq:bnd-overlap} the space of
boundary changing fields that separate a $(\Nr,\eta)$-boundary
condition from a $(\Dr,\eta)$-boundary condition forms the
representation $\Vc_{-\frac18}$ under the action of the triplet
algebra. To be specific, let us denote by 
$\mu^\eta(x)$ the boundary changing field that 
interpolates between the $(\Nr,\eta)$ boundary condition 
for boundary points $y < x$, and the $(\Dr,\eta)$ boundary condition
for boundary points $y > x$.  We also denote by  
$\tilde \mu^\eta(x)$ the boundary changing field which
conversely interpolates between  $(\Dr,\eta)$ for $y < x$ and 
$(\Nr,\eta)$ for $y > x$.
We begin by considering the OPE of $\mu^\eta(x)$ with $\om(0)$
to first subleading order. The representation $\Vc_{-\frac18}$ has
one-dimensional eigenspaces for the $L_0$ eigenvalues $-\frac18$ and
$-\frac18+1$, spanned by $\mu$ and $L_{-1}\mu$, respectively, so that
we can make the ansatz (for $x>0$)
\be
\mu^\eta(x)\,  \om(0) = 
f(x) \mu^\eta(0) + g(x) L_{-1} \mu^\eta(0) 
+ \orl[h{=}{-}\tfrac{1}{8}{+}2,x^2] \ ,
\labl{eq:muon-bnd-ope-ansatz}
where $f$ and $g$ are functions. As before, acting with $L_0$ on both
sides results in a differential equation for $f(x)$, namely this time
$xf'(x)+1=0$. Acting with $L_1$ on both sides gives in addition the 
condition 
\be
  \big( -\tfrac14 x + x^2 \tfrac\partial{\partial x} \big)
  \big( f(x) \mu(0) + g(x) L_{-1} \mu + \cdots \big)
  = g(x) L_1 L_{-1} \mu(0) + \orl[h{=}{-}\tfrac{1}{8}{+}1,x^2]\ ,
\ee
which shows that $g(x) = x(f(x){+}4)$. Thus altogether
\be
  \mu^\eta(x) \om(0) = \big( C_{\mu\om}^\eta - \log x\big) \mu^\eta(0) + 
  x\big( C_{\mu\om}^\eta +4 - \log x\big) L_{-1}\mu^\eta(0) 
+ \orl[h{=}{-}\tfrac{1}{8}{+}2,x^2]
  \ , 
\labl{eq:mu-om-ope-aux1}
for some constant $C_{\mu\om}^\eta$ to be determined below in a
crossing calculation.

As another ingredient in the crossing calculation we need two- and
three-point functions involving boundary changing fields. Let us write 
$\tilde\mu^\eta_\infty$ for the boundary field $\tilde\mu^\eta$ placed at
infinity on the upper half plane (see appendix \ref{app:local-coord}
for a more careful treatment of field insertions at infinity). 
Then $\langle \tilde\mu^\eta_\infty \, \mu^\eta(x) \rangle_\text{uhp}$
is independent of $x$ and gives the normalisation of the product 
$\tilde\mu^\eta \mu^\eta$. We will not fix this normalisation, but
rather leave the two-point correlator explicitly in intermediate
expressions. The three-point function is fixed by M\"obius covariance
up to constants. One finds
\be
  \corr{\, \tilde\mu^\eta_\infty \, \mu^\eta(x) \, 
\om(0) \,}_\text{uhp}
  = \big( C_{\mu\om}^\eta - \log x \big)
  \langle \tilde\mu^\eta_\infty \, \mu^\eta(0) \rangle_\text{uhp} \ ,
\labl{eq:mu-mu-om-bnd}
where the appearance of the OPE coefficient $C_{\mu\om}^\eta$ can be
deduced by comparing the small $x$ behaviour of \erf{eq:mu-mu-om-bnd}
with the OPE \erf{eq:mu-om-ope-aux1}.

We have now gathered all ingredients for the crossing calculation.
Consider the four point function
\be
  F(x) = \corr{\, \tilde\mu^\eta_\infty \, \mu^\eta(1) \, 
\om(x)\,\om(0) \,}_\text{uhp} \ , \qquad 0<x<1 \ .
\ee
The asymptotic expansions as given by the OPEs 
\erf{eq:omom-bnd-ope-aux1} and \erf{eq:mu-om-ope-aux1} are
\be
  F(\eps) \underset{\eps \rightarrow 0}=
  \big(C_\om^\eta -2 C_{\mu\om}^\eta \log\eps  
   - (\log\eps)^2 + \orl(\eps) \big) \cdot
  \langle \tilde\mu^\eta_\infty \, \mu^\eta(0) \rangle_\text{uhp}
\labl{eq:mmoo-bnd-asymp1}
and
\be
  F(1{-}\eps) \underset{\eps \rightarrow 0}=
  \big( (C_{\mu\om}^\eta)^2 - C_{\mu\om}^\eta \log\eps 
      - 4 \eps + \orl(\eps^2) \big) \cdot
  \langle \tilde\mu^\eta_\infty \, \mu^\eta(0) \rangle_\text{uhp} \ .
\labl{eq:mmoo-bnd-asymp2}
On the other hand, using a null vector argument outlined in
appendix \ref{app:mmoo-block}, $F(x)$ has to be equal to
\be
  F(x) = - C_1 h(x)^2 - C_2 \log(1{-}x) + C_4 h(x) + C_5  \ ,
  \qquad 
  h(x) = \log \left( \frac{ 1 - \sqrt{1-x}}{1+\sqrt{1-x}} \right) 
\ee
for an appropriate choice of constants $C_1$, $C_2$, $C_4$ and $C_5$.
Using the asymptotic behaviour \erf{eq:mmoo-bnd-asymp2} to fix these
constants results in
\be\begin{array}{ll}\displaystyle
  C_1 = \langle \tilde\mu^\eta_\infty \, \mu^\eta(0)
  \rangle_\text{uhp} \ ,    \qquad \etb
  C_2 = C_{\mu\om}^\eta \cdot
  \langle \tilde\mu^\eta_\infty \, \mu^\eta(0) \rangle_\text{uhp} \qquad
  \enl
  C_4 = 0 \ , \etb
  C_5 = (C_{\mu\om}^\eta)^2 \cdot
  \langle \tilde\mu^\eta_\infty \, \mu^\eta(0) \rangle_\text{uhp}\ .
\eear\ee
One can now compute the small $x$ behaviour of $F(x)$ to be
\be
  F(x) = \Big((C_{\mu\om}^\eta)^2 -(2 \log 2)^2 
+ 4 \log 2 \cdot \log x - (\log x)^2 + \orl(x) \Big)
\langle \tilde\mu^\eta_\infty \, \mu^\eta(0) \rangle_\text{uhp} \ .
\ee
Comparing to the asymptotic behaviour \erf{eq:mmoo-bnd-asymp1}
obtained using the OPE directly shows that there is a unique
consistent choice for the boundary OPE coefficients involved, namely 
\be
  C_\om^\eta = 0 \ , \qquad
  C_{\mu\om}^\eta = -2 \log 2 \ .
\labl{eq:mu-om-bndope-coeff}
This confirms the operator product expansion
\erf{eq:summary-omom-bnd-ope} quoted in the summary section. It also
shows that (at least to this order) the boundary fields define a
consistent (associative) operator product.\footnote{
After a short calculation using \erf{eq:om-bnd-2pt},
\erf{eq:mu-mu-om-bnd}  and \erf{eq:mu-om-bndope-coeff}
one finds for the boundary changing fields the two OPEs 
$\mu^\eta(x) \tilde\mu^\eta(0) = N_1 \, x^{\frac14}( \Omega(0) + \dots )$
and 
$\tilde\mu^\eta(x)\mu^\eta(0) 
= N_2 \,  x^{\frac14}( \omega(0) +  \log(x/4) \Omega(0) + \dots )$,
where $x>0$ and 
$N_1 = \langle \mu^\eta_\infty \tilde\mu^\eta(0) \rangle_\text{uhp}/
        \langle \one \rangle_\text{uhp}^{\Dr,\eta}$ and
$N_2 = \langle \tilde\mu^\eta_\infty \mu^\eta(0) \rangle_\text{uhp}/
       \langle \omega(0) \rangle_\text{uhp}^{\Nr,\eta}$ are constants
related to the normalisation of the fields, see also section
\ref{sec:b1-uhp} below. In the first case, a stretch of $(\Nr,\eta)$
boundary is collapsed and one is left with boundary fields on
$(\Dr,\eta)$; in the second case a stretch of $(\Dr,\eta)$ boundary
disappears and the resulting boundary fields live on the 
$(\Nr,\eta)$ boundary. However, we will not need these OPEs in the
present paper.} 

It should be pointed out that while the above arguments fix the
leading order in the operator product expansion
\erf{eq:omom-bnd-ope-aux1}  of two $\omega$ fields, in a more detailed
analysis one finds that there are still two undetermined constants at
subleading orders.  This is discussed in appendix \ref{app:V18-rep}.

\sect{Bulk-boundary OPE and factorisation}\label{sec:bbO+fac}

Next we want to determine the bulk-boundary operator product expansion
of the bulk fields $\bomega$ and $\bmu$ near a $(\Dr,\pm)$ and  
$(\Nr,\pm)$-boundary. This will allow us to check some of the
factorisation constraints, in particular the compatibility of our
boundary conditions with the bulk operator product expansion.

\subsection{Bulk-boundary OPE}\label{sec:bulk-bnd-ope}

Let $\bphi(iy)$ be a Virasoro primary bulk field in the upper half plane,
with $y>0$ the distance to the boundary. Depending on
whether we put the $(\Dr,\pm)$ or $(\Nr,\pm)$-boundary condition on the
real line, we can make the following  ansatz for the leading order in
the bulk-boundary operator product expansion, 
\bea
  \bphi(iy)\bnd{\Dr,\pm} =~ a(y) \Om(0) + \orl[h{=}2,y^2]
  \ , \enl
  \bphi(iy)\bnd{\Nr,\pm} =~ b(y) \Om(0) + c(y) \om(0) + \orl[h{=}1,y] \ , 
\eear\labl{eq:generic-bnd-OPE}
with $\orl[h,y]$ as defined in section \ref{sec:bndOPE-oo}.
Here $a(y), b(y), c(y)$ are functions that depend on the bulk field
$\bphi$ and the boundary condition. Note that the operator product
expansion for the $(\Dr,\pm)$-boundary condition does not contain a
linear $y$-term since the boundary fields transform in the 
representation $\Vc_0$ that does not have
a state at conformal weight one.  

To determine the functions $a(y), b(y), c(y)$, one first demands
consistency with the action of the Virasoro zero mode on the boundary,
$L_0^\text{bnd}$. Recall from \cite{Cardy:1989ir} that the Virasoro
modes acting on boundary fields on the upper half plane are build by
integrating $T(z)$ and $\ol T(z)$ on half-circles on the upper half
plane. Correspondingly, the commutation relation with the bulk-field
$\bphi(z)$ now reads 
\be
  [ L_m^\text{bnd} , \bphi(z) ]
  = \Big(\big( (m{+}1) z^m L_0 + z^{m+1} L_{-1} +
     (m{+}1) (z^*)^m \ol L_0 + (z^*)^{m+1} \ol L_{-1} \big) \bphi
	\Big)(z)  \ .   
\ee
In the case $z = iy$ and $m=0$ this simplifies to
\be
  [ L_0^\text{bnd} , \bphi(iy) ]
  = \big( (L_0 + \ol L_0)\bphi)(iy) + y \tfrac\partial{\partial y}
	\bphi(iy)  \ .   
\ee
Just as for the OPEs \erf{eq:om-bOPE-ansatz} and
\erf{eq:muon-bnd-ope-ansatz}, by acting with $L_0^\text{bnd}$ on both 
sides of \erf{eq:generic-bnd-OPE} we obtain differential equations for
$a(y), b(y), c(y)$. For the bulk fields $\bomega$ and $\bmu$ one
obtains in this way 
\be
\bearll
  \bomega(iy)\bnd{\Dr,\eta} \etb=~ ( C_\omega^{\Dr,\eta} -2 \log y) 
	\Omega(0)~+~ \orl[h{=}2,y^2]
  \enl
  \bomega(iy)\bnd{\Nr,\eta} \etb=~ D_\omega^{\Nr,\eta} \omega(0) +
    \big( (D_\omega^{\Nr,\eta}-2)\log y+ C_\omega^{\Nr,\eta}\big) 
    \Omega(0) ~+~ \orl[h{=}1,y]
  \enl
  \bmu(iy)\bnd{\Dr,\eta} \etb=~ C_\mu^{\Dr,\eta} (2y)^{\frac14}\big( 
    \Omega(0)  ~+~ \orl[h{=}2,y^2]\big)
  \enl
  \bmu(iy)\bnd{\Nr,\eta} \etb=~ D_\mu^{\Nr,\eta} (2y)^{\frac14} \big( 
     \omega(0) +
    (\log y+ C_\mu^{\Nr,\eta}) \Omega(0)  ~+~ \orl[h{=}1,y] \big)
\eear\labl{eq:bulkbndOPE-ansatz}
with a number of undetermined constants. Some of these are fixed by
comparing the OPEs to the one-point functions on the upper half plane
\erf{eq:bulk-1pt-uhp}. For example, using the first OPE in
\erf{eq:bulkbndOPE-ansatz} we get
\be
  \corr{\bomega(iy)}_{\rm uhp}^{{\rm D},\eta}
  = ( C_\omega^{\Dr,\eta} -2 \log y) \corr{\one}_{\rm uhp}^{
	{\rm D},\eta} 
    + \orl(y^2) 
  = -2 \log(2y) \corr{\one}_{\rm uhp}^{{\rm D},\eta} \ , 
\ee
where in the last equality the exact expression from
\erf{eq:bulk-1pt-uhp} was substituted. Here $\langle \one \rangle$
denotes the correlator with no field insertions. Note that an
insertion of the bulk vacuum $\bOmega(z)$ or the boundary vacuum
$\Omega(x)$ counts as no field insertion. In this way we arrive at 
\be
  C_\omega^{\Dr,\eta} = -2 \log 2 \ ,\qquad
  C_\mu^{\Dr,\eta} = \eta \sqrt{\pi} \ .
\labl{eq:bulk-bnd-D-consts}
This determines the bulk-boundary OPE of $\bomega$ and $\bmu$ close to
a $(\text{D},\eta)$-boundary. To find the coefficients for a
$(\text{N},\eta)$-boundary we perform a crossing calculation in the
presence of a boundary changing field. 

Consider a bulk field $\bphi(z)$ on the upper half plane, where on the
positive real axis we have the boundary condition $(\text{D},\eta)$,
and on the negative real axis the boundary condition
$(\text{N},\eta)$. At the points $0$ and $\infty$ boundary changing
fields have to be inserted, and the correlator we will consider is 
\be
  \corr{\,\tilde\mu_\infty^\eta\,\mu^\eta(0)\,\bphi(e^{i\theta}) 
 	}_{\text{uhp}} \ .
\labl{eq:bb-corr-interest}
The asymptotics of this block in the $\theta\rightarrow 0$ limit is
known from the OPE \erf{eq:bulkbndOPE-ansatz}, together with 
\erf{eq:bulk-bnd-D-consts}. Matching this with the four-point blocks
in appendix \ref{app:four-point-blocks} allows us to determine  
\erf{eq:bb-corr-interest} uniquely. Finally, evaluating the 
behaviour for $\theta\rightarrow \pi$ fixes the remaining constants in
\erf{eq:bulkbndOPE-ansatz}.

To carry out this calculation, we need to relate the correlator
\erf{eq:bb-corr-interest} to the four-point blocks with insertions at
$\infty$, $1$, $x$ and $0$ in appendix \ref{app:four-point-blocks}.
Suppose the bulk field $\bphi$ is of the form 
$\bphi = \phi_1 \otimes \phi_2 + \Nc$. Then the correlator
\erf{eq:bb-corr-interest} is equal to a conformal block 
with insertions 
$\langle \mu| \mu(1) \phi_1(e^{i\theta})\phi_2(e^{-i\theta})\rangle$.
The relevant M\"obius transformation is 
$\zeta \mapsto 1 - e^{i\theta} \zeta$, so that the correlator
\erf{eq:bb-corr-interest} is an element of the following space of
blocks, 
\be
  \corr{\,\tilde\mu_\infty^\eta\,\mu^\eta(0)\,\bphi(e^{i\theta}) 
	}_{\text{uhp}}  ~ \in ~
  \big\{\,
  \langle \mu| \mu(1) \tilde\phi_1(x)\tilde\phi_2(0)\rangle\,\big\}
  \ , \qquad   x = 1 - e^{2i\theta} \ ,
\labl{eq:bulk-bnd-corr-space}
where $\tilde\phi_1 = e^{i(\theta+\pi)L_0} \phi_1$ and
$\tilde\phi_2 = e^{i(\theta+\pi)L_0} \phi_2$.
Let us first consider the case $\bphi = \bmu$ and then go
on to $\bphi = \bomega$.

\subsubsection{The bulk field $\bmu$ in the presence of boundary
changing fields} 

Substituting the general solution for the four-point block with
four $\mu$-insertions from \erf{eq:mmmm-space} into 
\erf{eq:bulk-bnd-corr-space} gives
\be
  \corr{\,\tilde\mu_\infty^\eta\,\mu^\eta(0)\,\bmu(e^{i\theta}) 
	}_{\text{uhp}}
  = \big(2 \sin\theta \big)^{\frac14} e^{i \theta/2}
  \big( C_1 F(x) + C_2 G(x) \big)
  \qquad \text{for~some}~~ C_1,C_2 \in \Cb \ ,
\ee
where as before $x = 1-e^{2i\theta}$ and furthermore
\bea
  F(x) = {}_2 F_1(\tfrac{1}{2},\tfrac{1}{2};1;x) 
	= 1 + \tfrac14 x + \Oc(x^2) \ ,
  \enl
  G(x) = {}_2 F_1(\tfrac{1}{2},\tfrac{1}{2};1;1{-}x) = 
  \frac{1}{\pi}\Big( 4 \log2 - \log x + 
	( \log2 - \tfrac12 - \tfrac14\log x) x  + \orl(x^2) \Big) \ .
\eear\labl{eq:FG-hyper}
Using the OPE \erf{eq:bulkbndOPE-ansatz} close to a
$(\text{D},\eta)$-boundary gives the asymptotic behaviour
\be
  \corr{\,\tilde\mu_\infty^\eta\,\mu^\eta(0)\,\bmu(e^{i\theta}) 
	}_{\text{uhp}}
  = \eta \sqrt{\pi} (2\theta)^{\frac14}\Big( 
  \corr{\tilde\mu_\infty^\eta\,\mu^\eta(0)}_{\text{uhp}}
  + \orl(\theta^2)\Big) \ . 
\ee
This fixes $C_1 = \eta \sqrt{\pi} 
\langle \tilde\mu_\infty^\eta\mu^\eta(0)\rangle_{\text{uhp}}$ and $C_2=0$,
so that altogether the correlator is equal to
\be
  \corr{\,\tilde\mu_\infty^\eta\,\mu^\eta(0)\,\bmu(e^{i\theta}) 
	}_{\text{uhp}}
  = \eta \sqrt{\pi} \, \big(2 \sin\theta \big)^{\frac14} e^{i \theta/2}
  {}_2 F_1(\tfrac{1}{2},\tfrac{1}{2};1;1{-}e^{2i\theta})
  \, \corr{\tilde\mu_\infty^\eta\,\mu^\eta(0)}_{\text{uhp}} \ .
\labl{eq:bnd-ch-mu}
Next we need to evaluate the asymptotic behaviour for 
$\theta \rightarrow \pi$. To do so, first note that varying the angle
$\theta$ from $0$ to $\pi$ amounts to moving the point 
$x = 1-e^{2i\theta}$ counter-clockwise around 
the point $1$. Since $1$ is a branchpoint of $F(x)$, we have to
analytically continue $F(x)$ before evaluating the $x \rightarrow 0$
behaviour (which amounts to $\theta \rightarrow \pi$). Let $\Cc_1$ be
the analytic continuation of the argument $x$ counter-clockwise around
the point $1$. For the corresponding monodromy one finds
\be
  F(x) \overset{\Cc_1}{\longrightarrow} F(x) - 2 i G(x)
  \ , \qquad
  G(x)  \overset{\Cc_1}{\longrightarrow} G(x) \ .
\ee
The asymptotics of \erf{eq:bnd-ch-mu} is then given by
\bea
  \corr{\,\tilde\mu_\infty^\eta\,\mu^\eta(0)\,\bmu(e^{i(\pi-\eps)}) 
	}_{\text{uhp}}
  \enl
  = \eta \sqrt{\pi} \, \big(2 \sin\eps \big)^{\frac14} i e^{-i \eps/2}
  \big( F(1{-}e^{-2i\eps}) - 2 i G(1{-}e^{-2i\eps}) \big)
  \, \corr{\tilde\mu_\infty^\eta\,\mu^\eta(0)}_{\text{uhp}}
  \enl
  = \eta \sqrt{\pi} \, (2 \eps)^{\frac14} 
  \big( \tfrac6\pi \log2 - \tfrac2\pi \log\eps + \orl(\eps)\big)
  \, \corr{\tilde\mu_\infty^\eta\,\mu^\eta(0)}_{\text{uhp}} \ .
\eear\ee
This has to be compared with the asymptotics obtained using the OPE
\erf{eq:bulkbndOPE-ansatz}
of $\bmu$ close to a $(\text{N},\eta)$-boundary, which is
\be
  \corr{\,\tilde\mu_\infty^\eta\,\mu^\eta(0)\,\bmu(e^{i(\pi-\eps)}) 
	}_{\text{uhp}}
  = D_\mu^{\Nr,\eta} (2\eps)^{\frac14} \big( C_\mu^{\Nr,\eta}  
    - 2\log2 + \log\eps + \orl(\eps)\big)
  \, \corr{\tilde\mu_\infty^\eta\,\mu^\eta(0)}_{\text{uhp}} \ ,
\ee
where we also used the boundary three point function
\erf{eq:mu-mu-om-bnd} and the structure constant
\erf{eq:mu-om-bndope-coeff}. The resulting values of the so far
undetermined constants are now fixed uniquely to be 
\be
  D_\mu^{\Nr,\eta} = - 2\eta \, \pi^{-\frac12} \ , \qquad
  C_\mu^{\Nr,\eta} = -\log 2 \ .
\ee
Altogether, for $\bmu$ we obtain the bulk-boundary OPE quoted in
\erf{eq:bulk-bnd-OPE-result} in the introduction.

\subsubsection{The bulk field $\bomega$ in the 
presence of boundary changing fields}

For $\bphi = \bomega = \omega \otimes \omega + \Nc$ we need the
space of blocks \erf{eq:bulk-bnd-corr-space} with
$\tilde\phi_1 = \tilde\phi_2 = \omega + i(\theta{+}\pi)\Omega$.
The relevant individual four-point blocks are given in
\erf{eq:mmoo-Om-ins} and \erf{eq:mmoo-gx}. This results in the
following ansatz for the correlator
\bea
  \corr{\,\tilde\mu_\infty^\eta\,\mu^\eta(0)\,\bomega(e^{i\theta}) 
	}_{\text{uhp}}
  \enl
  = - C_1 h(x)^2 + C_4 h(x) + C_1 \theta^2 + i(C_3{-}C_2) \theta 
    + C_5 + i \pi (C_2{+}C_3) - \pi^2 C_1  
  \enl
  = -\big(\pi^2 + (\tfrac{i \pi}2{+}\log 2)^2 \big)C_1
  + i \pi(C_3{+}C_2) - (\tfrac{i \pi}2{+}\log 2) C_4 + C_5
  \enl
  \qquad
    + (C_4+2(\tfrac{i \pi}2{+}\log 2)C_1) \log\theta
    - C_1 (\log\theta)^2
    + i (C_3{-}C_2) \theta + \orl(\theta^2) \ .
\eear\labl{eq:bulk-bnd-corr-space-om}
Here $C_1,\dots,C_5 \in \Cb$ are constants (used in the same way
as in \erf{eq:mmoo-Om-ins} and \erf{eq:mmoo-gx}), the function
$h(x)$ is given in \erf{eq:mmoo-gx}, and as before 
$x = 1{-}e^{2i\theta}$. 
Using the OPE of $\bomega$ close to a $(\text{D},\eta)$-boundary we
obtain 
\be
  \corr{\,\tilde\mu_\infty^\eta\,\mu^\eta(0)\,\bomega(e^{i\theta}) 
	}_{\text{uhp}}
  = -2 \log(2\theta) \, \corr{\tilde\mu_\infty^\eta\,\mu^\eta(0)
	}_{\text{uhp}}     + \orl(\theta^2) \ .
\ee
Comparing to \erf{eq:bulk-bnd-corr-space-om} it follows that 
$C_1=0$, $C_2-C_3=0$, {\it etc}. Altogether, the correlator is simply 
\be
  \corr{\,\tilde\mu_\infty^\eta\,\mu^\eta(0)\,\bomega(e^{i\theta}) 
	}_{\text{uhp}}
  = -2 \log\!\big( 4 \tan\tfrac{\theta}{2} \big)
  \cdot \corr{\tilde\mu_\infty^\eta\,\mu^\eta(0)}_{\text{uhp}}  \ .
\ee
For $\theta = \pi{-}\eps$ and $\eps \rightarrow 0$ 
this correlator behaves as $(-6\log 2+2\log\eps + \orl(\eps)) 
\langle \tilde\mu_\infty^\eta\mu^\eta(0)\rangle_{\text{uhp}}$, which has
to be matched against the expression obtained using the OPE
close to the $(\text{N},\eta)$ boundary,
\bea
  \corr{\,\tilde\mu_\infty^\eta\,\mu^\eta(0)\,\bomega(e^{i(\pi-\eps)}) 
	}_{\text{uhp}}
  \enl
  = \big( -2 \log 2 \cdot D_\omega^{\Nr,\eta} + C_\omega^{\Nr,\eta} + 
    (D_\omega^{\Nr,\eta}{-}2)\log \eps + \orl(\eps) \big)
    \corr{\tilde\mu_\infty^\eta\,\mu^\eta(0)}_{\text{uhp}} \ ,
\eear\ee
where as before we also used \erf{eq:mu-mu-om-bnd} and 
\erf{eq:mu-om-bndope-coeff}. The resulting constants are
\be
  D_\omega^{\Nr,\eta} = 4 \ , \qquad
  C_\omega^{\Nr,\eta} = 2 \log 2 \ ,
\ee
which when inserted into \erf{eq:bulkbndOPE-ansatz} indeed gives 
\erf{eq:bulk-bnd-OPE-result}.

\subsubsection{Bulk one-point correlators on the upper half plane}
\label{sec:b1-uhp}

Having established the list of OPEs in \erf{eq:bulk-bnd-OPE-result},
we note that 
$\langle \bomega(iy) \rangle^{\Nr,\eta}_\text{uhp} = 
4 \langle \omega(0) \rangle^{\Nr,\eta}_\text{uhp}$. Together
with \erf{eq:bulk-1pt-uhp} we therefore have 
\be
  \corr{\one}^{\Dr,\eta}_\text{uhp} = \tfrac12 \lambda \eta  \pi^{-\frac12}
  \qquad \text{and} \qquad
  \corr{ \omega(0) }^{\Nr,\eta}_\text{uhp} = 
  -\tfrac12 \lambda \eta \pi^{\frac12}   \ .
\ee
These are the simplest non-zero correlators on the upper half plane
with $(\Dr,\eta)$ and $(\Nr,\eta)$ boundary condition, respectively,
and we will normalise other uhp-correlators relative to these. For
example, for the bulk one-point functions in \erf{eq:bulk-1pt-uhp} 
we get
\be
\bearll
  \corr{\bomega(iy)}_{\rm uhp}^{{\rm D},\eta} = 
   - 2 \log(2 y) \corr{\one}^{\Dr,\eta}_\text{uhp} \ ,
  \quad \etb  \quad 
  \corr{\bomega(iy)}_{\rm uhp}^{{\rm N},\eta} = 
        4 \corr{ \omega(0) }^{\Nr,\eta}_\text{uhp} \ ,
  \enl
  \corr{\bmu(iy)}_{\rm uhp}^{{\rm D},\eta} = 
   \eta \pi^{\frac12} \, (2y)^{\frac{1}{4}}
       \corr{\one}^{\Dr,\eta}_\text{uhp} \ ,
  \etb \quad
  \corr{\bmu(iy)}_{\rm uhp}^{{\rm N},\eta} = 
      -2 \eta \pi^{-\frac12} \,  (2y)^{\frac{1}{4}}
      \corr{ \omega(0) }^{\Nr,\eta}_\text{uhp} \ .
\eear\labl{eq:bulk-1pt-uhp-norm}

\subsection{Factorisation for two bulk fields on the upper half plane}

The computation of the bulk-boundary OPE \erf{eq:bulk-bnd-OPE-result}
did use implicitly the {\em bulk spectrum} of the triplet theory,
which entered the process of finding a consistent ansatz for the
boundary states. However, we should stress that the {\em bulk OPEs} 
\erf{eq:bulk-OPE-result} have not been used at any point of the
calculation so far. These bulk OPEs were determined in
\cite{Gaberdiel:1998ps} by demanding crossing symmetry for correlators
on the complex plane without reference to any boundary conditions.

By performing a crossing calculation with the correlator of two bulk
fields on the upper half plane, which links the limit where the two
bulk fields approach the boundary to the limit where the two bulk
fields approach each other, we can compare the bulk OPE to the
bulk-boundary OPE coefficients.  This provides a significant 
consistency check (which is sometimes called the cluster condition or
classifying algebra) on the bulk-boundary OPE and hence also on the
boundary field content found in the boundary state analysis. In
\cite{Cardy:1991tv} the opposite procedure was used to deduce the
bulk-boundary OPE from the known bulk OPE. 
\medskip

Consider thus the correlator of two bulk fields 
$\bphi = \phi_1 \otimes \phi_2 + \Nc$ and
$\bpsi = \psi_1 \otimes \psi_2 + \Nc$ on the upper half plane with
boundary condition $b$. The chiral Ward identities demand this
correlator to be an element of the following space of 
conformal four-point blocks\footnote{This is what is sometimes
referred to as the doubling trick.} 
\be
  \corr{ \bphi(x{+}iy)\,\bpsi(iy)}^b_{\text{uhp}}
  ~ \in ~
  \big\{\,
  \langle \phi_1(x{+}iy) \phi_2(x{-}iy) \psi_1(iy) \psi_2({-}iy)
  \rangle\,\big\} \ ,
\labl{eq:bulk-bnd-corr-space1}
where $x,y \in \Rb_{>0}$. To make use of the four-point blocks
derived in appendix \ref{app:four-point-blocks} 
one has to apply a M\"obius transformation
to move the insertion points to $\infty$, $1$, $r$ and $0$. We
will use
\be
  M(\zeta) = \frac{2iy}{2iy-x} \cdot \frac{\zeta - iy}
	{\zeta - (x{+}iy)} \ ,
\ee
which takes $x{+}iy$ to infinity, {\it etc}. The scaling factors
arising in this transformation are easiest to determine by explicitly
working with the local coordinates around the insertion points as in  
appendix \ref{app:local-coord}. One finds that the uhp-correlator is
equally an element of the following space of blocks,
\be
  \corr{ \bphi(x{+}iy)\,\bpsi(iy)}^b_{\text{uhp}}
  ~ \in ~
  \big\{\,
  \langle \tilde\phi_1 | \tilde\phi_2(1) 
  \tilde\psi_2(r) \tilde\psi_1(0) \rangle\,\big\} \ , 
  \quad \text{where} \quad r = \frac{(2y)^2}{x^2+(2y)^2}
\labl{eq:2bulk-uhp-blocks}
and
\be\begin{array}{llll}\displaystyle
  \tilde\phi_1 = (f_\infty)^{L_0} \phi_1 ~,~\etb
  \tilde\phi_2 = (f_1)^{L_0} \, \phi_2      ~,~\etb
  \tilde\psi_2 = (f_r)^{L_0} \, \psi_2      ~,~\etb
  \tilde\psi_1 = (f_0)^{L_0} \, \psi_1       \ , 
  \enl
  f_\infty = \frac{x{-}2iy}{2ixy}       ~,\etb
  f_1 = \frac{x}{2iy(x{-}2iy)}           ~,\etb
  f_r = \frac{2ixy}{(x{-}2iy)(x{+}2iy)^2}\,,\etb
  f_0 = \frac{2iy}{x(x{-}2iy)} \ .
\eear\labl{eq:2bulk-uhp-details}
We will present the calculations for the three correlators
$\langle \bmu(z) \bmu(w) \rangle_\text{uhp}$, 
$\langle \bmu(z) \bomega(w) \rangle_\text{uhp}$, and
$\langle \bomega(z) \bomega(w) \rangle_\text{uhp}$, in this order.

\subsubsection{Two $\bmu$ fields on the upper
half plane}\label{sec:fact-mumu-uhp}

In this case one has to apply \erf{eq:2bulk-uhp-blocks} to the
case $\bphi = \bpsi = \mu \otimes \mu + \Nc$. Substituting also
the general form \erf{eq:mmmm-space} of the four-point blocks, a short
calculation yields
\bea
  \corr{ \bmu(x{+}iy)\,\bmu(iy)}^b_{\text{uhp}}
  = \Bigg( \frac{(2xy)^2}{x^2 + (2y)^2} \Bigg)^{\frac14}
    \big( A_1 F(r) + A_2 G(r) \big)
  \enl
  \qquad \overset{y\rightarrow 0}= 
  (2y)^{\frac12}\big( A_1 +  A_2 \tfrac{2}{\pi} \log 2 \,+\,
    A_2  \tfrac{2}{\pi} \log \tfrac{x}{y} ~ + \orl(y^2) \big)
  \enl
  \qquad \overset{x\rightarrow 0}= x^{\frac12} \big( 
    A_1 \tfrac{6}{\pi} \log 2 + A_2 \,-\, A_1  
	\tfrac{2}{\pi} \log \tfrac{x}{y}
    ~+ \orl(x^2) \big)
\eear\labl{eq:mumu-uhp-space}
with $F$ and $G$ as in \erf{eq:FG-hyper} and $A_1,A_2 \in \Cb$ constants
to be determined. The $y\rightarrow 0$ behaviour is fixed by the
bulk-boundary OPE \erf{eq:bulk-bnd-OPE-result},
\be
    \corr{ \bmu(x{+}iy)\,\bmu(iy)}^b_{\text{uhp}}
  = (2y)^{\frac12} \cdot \begin{cases}
  \pi \langle \one \rangle_\text{uhp}^{\text{D},\eta}  + \orl(y^2) & 
    ;~b = \text{D},\eta \\[.5em]
  -\tfrac{8}{\pi}\big( \log 2 + \log \tfrac{x}{y} \big)
    \langle \omega(0) \rangle_\text{uhp}^{\text{N},\eta} + \orl(y^2)
     & ;~b = \text{N},\eta \ . 
  \end{cases}
\ee
For the $(\text{N},\eta)$-boundary we also used \erf{eq:om-bnd-2pt} and 
the fact that $\langle \one \rangle_\text{uhp}^{\text{N},\eta} = 0$
(cf.~\erf{eq:bulk-1pt-uhp}) and that 
$\langle \omega(x) \, (W^a_{-1} \omega)(y) 
\rangle_{\rm uhp}^{{\rm N},\eta}=0$. Comparing to
\erf{eq:mumu-uhp-space} yields
\be\begin{array}{lll}\displaystyle
  b = \text{D},\eta \etb : \quad 
  A_1 = \pi \langle \one \rangle_\text{uhp}^{\text{D},\eta}\ ,   \etb~ 
	A_2 = 0 \ ,
  \enl
  b = \text{N},\eta \etb : \quad A_1 = 0 \ , \etb~ 
  A_2 = -4\langle \omega(0) \rangle_\text{uhp}^{\text{N},\eta} \ .
\eear\labl{eq:mumu-uhp-A1A2}
By substituting these results into \erf{eq:mumu-uhp-space} we can then
evaluate the $x\rightarrow 0$ behaviour. 

Alternatively, this behaviour
can be obtained using the bulk OPE \erf{eq:bulk-OPE-result}. 
Together with \erf{eq:bulk-1pt-uhp-norm} one obtains
\be
    \corr{ \bmu(x{+}iy)\,\bmu(iy)}^b_{\text{uhp}}
  = x^{\frac12} \cdot \begin{cases}
     \big(6 \log 2 - 2 \log \tfrac{x}{y} \big)
     \langle \one \rangle_\text{uhp}^{\text{D},\eta} + \orl(x^2) & 
    ;~b = \text{D},\eta \\[.5em]
   -4\langle \omega(0) \rangle_\text{uhp}^{\text{N},\eta} + \orl(x^2)
     & ;~b = \text{N},\eta
  \end{cases}
\ee
in perfect agreement with \erf{eq:mumu-uhp-space} and 
\erf{eq:mumu-uhp-A1A2}.

\subsubsection{One $\bmu$ and one $\bomega$ field on the 
upper half plane}\label{sec:fact-muom-uhp}

For this calculation we will set $\phi_1=\phi_2=\mu$ and
allow $\psi_1,\psi_2 \in \{ \omega,\Omega \}$. Let us abbreviate
a conformal four-point block with two $\mu$-insertions and 
two arbitrary insertions $\alpha$, $\beta$ as
\be
  U(\alpha,\beta) = 
  \langle \mu | \mu(1) \alpha(r) \beta(0) \rangle \ .
\ee
Combining \erf{eq:2bulk-uhp-blocks} and \erf{eq:2bulk-uhp-details}
results in the following expressions for the bulk two-point
function on the upper half plane,
\bea
  \corr{ \bmu(x{+}iy)\,\bpsi(iy) }^b_\text{uhp}
  = (2y)^{\frac14}\Big(
  U(\psi_2,\psi_1) + \log f_r \cdot U(L_0\psi_2,\psi_1) 
\enl
\hspace*{10em}
  + \log f_0 \cdot U(\psi_2,L_0\psi_1) 
  + \log f_r \cdot \log f_0 \cdot U(L_0\psi_2,L_0\psi_1) \Big)
  \ .
\eear\ee
The relevant conformal blocks are given in
\erf{eq:mmoo-gdef}, \erf{eq:mmoo-Om-ins} and \erf{eq:mmoo-gx} 
in terms of five free parameters $C_1,\dots,C_5$.

To fix some of these parameters, let us first choose
$\psi_1=\psi_2=\Omega$, {\it i.e.}\ set 
$\bpsi = N = \Omega \otimes \Omega + \Nc$.  
Since $\Omega \otimes \Omega \in \Nc$, we have $N=0$
in the quotient space $\Hc^\text{bulk}$, and a correlator involving an
insertion of $N$ has to vanish. In particular,
\be
  0 = \corr{ \bmu(x{+}iy)\,N(iy) }^b_\text{uhp}
  = (2y)^{\frac14} U(\Omega,\Omega) = (2y)^{\frac14} C_1 \ ,
\ee
which implies that $C_1=0$, independent of the choice of boundary
condition $b$. Similarly, we have
$\omega \otimes \Omega - \Omega \otimes \omega \in \Nc$. If we
set $V_l = \omega \otimes \Omega + \Nc$ and
$V_r = \Omega \otimes \omega + \Nc$, then $V_l=V_r$ in 
$\Hc^\text{bulk}$ and hence
\bea
  0 = \corr{ \bmu(x{+}iy)\,V_l(iy) }^b_\text{uhp}
      - \corr{ \bmu(x{+}iy)\,V_r(iy) }^b_\text{uhp}
  \enl   \qquad 
  = (2y)^{\frac14}\big(
  U(\Omega,\omega) - U(\omega,\Omega) \big)
  = (2y)^{\frac14}\big( C_2 - C_3 \big) \ ,
\eear\ee
where in the last step we used that $C_1=0$. We thus get the
additional condition $C_2=C_3$. Altogether, the functional form
of the two-point function of $\bmu$ and $\bomega$ is now
restricted to
\bea
  \corr{ \bmu(x{+}iy)\,\bomega(iy) }^b_\text{uhp}
  = (2y)^{\frac14}\big( C_2(\pi i+ \log r - 2\log x) 
		+ C_4 h(r) + C_5 \big)
\enl
  \quad \overset{y\rightarrow 0}= 
  (2y)^{\frac14}\big(
  C_2(i\pi{+}2\log2) + C_5 -2(2C_2{+}C_4)\log x + 2(C_2{+}C_4)\log y
    ~ + \orl(y^2) \big)
  \enl
  \quad \overset{x\rightarrow 0}= (2y)^{\frac14}\big(
  i \pi C_2 + C_5 - 2 C_2 \log x ~+ \orl(x) \big) \ .
\eear\labl{eq:mu-om-blocklimits}
To determine the remaining constants we need to compare to the  
$y\rightarrow 0$ behaviour obtained by the
bulk-boundary OPE \erf{eq:bulk-bnd-OPE-result},
\bea
    \corr{ \bmu(x{+}iy)\,\bomega(iy)}^b_{\text{uhp}}
\enl
  = (2y)^{\frac14} \cdot \begin{cases}
  - 2 \eta \pi^{\frac12} \log(2y) 
  \langle \one \rangle_\text{uhp}^{\text{D},\eta}  + \orl(y) & 
    ;~b = \text{D},\eta \\[.5em]
  2 \eta \pi^{-\frac12} \big( 8 \log x - 6 \log y +
  2 \log 2 \big) 
  \langle \omega(0) \rangle_\text{uhp}^{\text{N},\eta} 
   + \orl(y)
     & ;~b = \text{N},\eta \ . 
  \end{cases}
\eear\ee
This fixes $C_2$, $C_4$ and $C_5$ to be
\be\begin{array}{llll}\displaystyle
  b = \text{D},\eta \etb : \quad 
  C_2 = \eta \sqrt{\pi} \langle \one \rangle_\text{uhp}^{\text{D},\eta}
	\ ,   \etb ~   C_4 = -2 C_2 \ ,  \etb~
  C_5 = -(i\pi + 4 \log 2) C_2 \ , 
  \enl
  b = \text{N},\eta \etb : \quad 
  C_2 = - \frac{2\eta}{\sqrt{\pi}} 
  \langle \omega(0) \rangle_\text{uhp}^{\text{N},\eta} \ ,
  \etb~   C_4 = 2 C_2 \ ,   \etb~  C_5 = -(i\pi + 4 \log 2) C_2 \ . 
\eear\labl{eq:mumu-uhp-C2C5}
Note that substituting the expression for $C_5$ simplifies the 
$x\rightarrow 0$ asymptotics in \erf{eq:mu-om-blocklimits} to
$-2(2\log2+\log x) (2y)^{\frac14} C_2 + \orl(x)$.
Next we compute the $x\rightarrow 0$ behaviour from the
bulk OPE \erf{eq:bulk-OPE-result}. One obtains
\bea
  \corr{ \bmu(x{+}iy)\,\bomega(iy)}^b_{\text{uhp}}
  \enl
  = -2(2\log 2 + \log x) \cdot  
  (2y)^{\frac14} \cdot \begin{cases}
  \eta \pi^{\frac12}
  \langle \one \rangle_\text{uhp}^{\text{D},\eta}  + \orl(x) & 
    ;~b = \text{D},\eta \\[.5em]
  -2 \eta \pi^{-\frac12} 
  \langle \omega(0) \rangle_\text{uhp}^{\text{N},\eta} 
   + \orl(x)
     & ;~b = \text{N},\eta \ . 
  \end{cases}
\eear\ee
Again, we find perfect agreement with
\erf{eq:mu-om-blocklimits} and \erf{eq:mumu-uhp-C2C5}.

\subsubsection{Two $\bomega$ fields on the upper half plane}

The calculation for two $\bomega$ fields on the upper half plane is
more involved because of the large number of free parameters in a four
point block with all insertions from $\Rc_0$. 
The relevant blocks are obtained by applying the
projection in appendix \ref{app:boson-blocks} to the fermionic blocks
found in appendix \ref{app:oooo-block}. 
The free parameters can be fixed by requiring 
first of all that the conformal
block gives zero if in \erf{eq:2bulk-uhp-blocks} one chooses one of
$\bphi$ or $\bpsi$ to be $\Omega \otimes \Omega + \Nc$ or 
$\omega \otimes \Omega - \Omega \otimes \omega + \Nc$.
Second, one has to match the functional form dictated by the
conformal block against the asymptotic behaviour determined
by the bulk-boundary OPE,
\be
    \corr{ \bomega(x{+}iy)\,\bomega(iy)}^b_{\text{uhp}}
  = \begin{cases}
  \big( 2\log (2y) \big)^2 
  \langle \one \rangle_\text{uhp}^{\text{D},\eta}  + \orl(y^2) & 
    ;~b = \text{D},\eta \\[.5em]
  16(\log(2y)-2\log x)
    \langle \omega(0) \rangle_\text{uhp}^{\text{N},\eta} + \orl(y)
     & ;~b = \text{N},\eta \ .
  \end{cases}
\labl{eq:omom-bnd-asymp-y=0}
It turns out that the leading order of the $y\rightarrow 0$
asymptotics as given above is not enough to determine all free parameters
of the relevant conformal block. The remaining constants can in principle
be fixed by calculating subleading orders of the OPE. However, we will
not do this and instead use the $x \rightarrow 0$ behaviour,
{\it i.e.}\ the bulk OPE to fix the remaining constants
\be
    \corr{ \bomega(x{+}iy)\,\bomega(iy)}^b_{\text{uhp}}
  = \begin{cases}
  4 \log x \cdot \big( 2 \log(2y)-\log x\big)
  \langle \one \rangle_\text{uhp}^{\text{D},\eta} 
  + \orl(x)~~, & 
    ;~b = \text{D},\eta \\[.5em]
  -16 \log x \cdot
    \langle \omega(0) \rangle_\text{uhp}^{\text{N},\eta} + \orl(x)
     & ;~b = \text{N},\eta \ .
  \end{cases}
\labl{eq:omom-bnd-asymp-x=0}
Of course, explicitly using the bulk OPE to fix the parameters in the
conformal blocks provides a less strong consistency check than 
the calculation in sections \ref{sec:fact-mumu-uhp} and 
\ref{sec:fact-muom-uhp}. 
Nonetheless, it is 
nontrivial that a solution with the correct limits exists.

We carried out the computations outlined above 
with the help of computer algebra
and merely state the result one finds for the correlator, 
\be
    \corr{ \bomega(x{+}iy)\,\bomega(iy)}^b_{\text{uhp}}
  = \begin{cases}
  \Big(
  \big( 2\log(2y) \big)^2 - \big( \log\big( 1{+}(2y/x)^2 \big)\big)^2
  \Big) \langle \one \rangle_\text{uhp}^{\text{D},\eta} 
  & ;~b = \text{D},\eta \\[.5em] \displaystyle
 8 \log\!\Bigg( \frac{(2y/x)^2}{x^2{+}(2y)^2} \Bigg)
    \langle \omega(0) \rangle_\text{uhp}^{\text{N},\eta}
     & ;~b = \text{N},\eta \ .
  \end{cases}
\ee
It is easy to see that this reproduces \erf{eq:omom-bnd-asymp-y=0}
and \erf{eq:omom-bnd-asymp-x=0}, while it requires a bit of work to
verify that it is in the correct space of conformal blocks, and that
it is indeed uniquely fixed by imposing the asymptotics
\erf{eq:omom-bnd-asymp-y=0} and \erf{eq:omom-bnd-asymp-x=0}.

\section{Conclusions}

In this paper we have constructed the boundary theory for the
logarithmic triplet theory at $c=-2$ in detail.
We started from the four boundary conditions compatible with the
free fermion symmetry, and proved that no additional boundary
conditions arise when only the triplet algebra is preserved.
These four boundary conditions can be labelled by the four
irreducible highest weight representations of the triplet algebra, and
the corresponding open string multiplicities are precisely described
by the fusion rules. The open string representations therefore include
also indecomposable (but reducible) representations of the triplet
algebra. 

We have calculated the OPE coefficients of the most
interesting boundary operators, namely those which are highest
weight with respect to the fermion modes; 
we have also determined the leading
bulk-boundary OPE coefficients. We have found consistent
solutions to all factorisation constraints that we have checked. We
regard this as very convincing evidence that we have found
a consistent boundary theory for the triplet model. 

One may expect that a similar approach should also be possible for 
other `rational' logarithmic theories, for example for the whole
family of $(1,q)$ triplet theories. It would be interesting to
understand which features of our analysis generalise directly to these
other cases. 
\smallskip

One of the motivations of this work was to analyse to which extent
logarithmic theories fit into the framework that was developed in 
\cite{Fuchs:2002cm,Fuchs:2004xi,Fjelstad:2005ua}
for (non-logarithmic) rational conformal field theories.
There it is shown that a consistent local conformal field theory
can be constructed starting from the algebra of boundary fields for
a single boundary condition. Knowledge of the bulk theory or of the
other boundary conditions is not required in this construction.
Here we have used a `hybrid' approach: the bulk-boundary OPE
coefficients were obtained from the algebra of boundary fields, but
the derivation made use of the bulk and boundary spectra since we used
the boundary states. It would obviously be interesting to understand
whether the above programme can also be applied to this theory. 
In particular, is it possible to obtain the entire triplet theory
starting with the (trivial) algebra structure of the $({\rm D},\pm)$
boundary that only has the vacuum representation $\Vc_0$ in its
spectrum? (In the rational case, the vacuum representation gives rise
to the `Cardy case', {\it i.e.}\ the conformal field theory with
charge conjugation modular invariant \cite{Fuchs:2002cm}.)

While the algebra on the $({\rm D},\pm)$ boundary is unique, we found
in appendix \ref{app:alg-assoc} a family of
non-isomorphic algebras on the representation $\Rc_0$ that makes up
the spectrum of the $({\rm N},\pm)$ boundary. This is 
something which cannot happen in the rational case. Only one of these
algebras was relevant for the triplet theory; it would be interesting
to understand the significance of these other algebras. We hope to
return to these points in the future.

\subsection*{Acknowledgements}

We thank J\"urgen Fuchs,
Thomas Quella, Philippe Ruelle and G\'erard Watts for
conversations and encouragement. The research of MRG was
partially supported by the Swiss National Science Foundation and the
Marie Curie network `Constituents, Fundamental Forces and Symmetries
of the Universe' (MRTN-CT-2004-005104). The research of IR 
was partially supported by the EU Research Training Network grants 
`Euclid', contract number HPRN-CT-2002-00325, and `Superstring Theory', 
contract number MRTN-CT-2004-512194.

\appendix

\sect{The triplet algebra and its structure constants}\label{app:trip}

The commutation relations of the triplet algebra are 
\begin{eqnarray}
  {}[ L_m, L_n ] &=& (m-n)L_{m+n} - \frac16 m(m^2-1) \delta_{m+n}
  \nonumber \\
  {}[ L_m, W^a_n ] &=& (2m-n) W^a_{m+n}
  \nonumber \\
  {}[ W^a_m, W^b_n ] &=& g^{ab} \biggl( 
  2(m-n) \Lambda_{m+n} 
  +\frac{1}{20} (m-n)(2m^2+2n^2-mn-8) L_{m+n} 
  \nonumber\\&&\qquad
  -\frac{1}{120} m(m^2-1)(m^2-4)\delta_{m+n}
  \biggr) 
   \\&&
  + f^{ab}_c \left( \frac{5}{14} (2m^2+2n^2-3mn-4)W^c_{m+n} 
    + \frac{12}{5} V^c_{m+n} \right)\ . \nonumber
  \end{eqnarray}
Here $\Lambda = \mathopen:L^2\mathclose: - 3/10\, \partial^2L$ and 
$V^a = \mathopen:LW^a\mathclose: - 3/14\, \partial^2W^a$ are
quasiprimary normal ordered fields. $g^{ab}$ and $f^{ab}_c$ are the
metric and structure constants of $su(2)$. In an orthonormal basis we
have $g^{ab} = \delta^{ab}, f^{ab}_c = i\epsilon^{abc}$; for the usual
Cartan-Weyl basis that we shall mainly use in this paper, the
conventions are  
\begin{equation}
f^{0\pm}_\pm = \pm1\ , \qquad f^{\pm\mp}_0 = \pm2\ , \qquad
\hbox{and} \qquad  
g^{00} = 1\ , \qquad g^{\pm\mp} = 2\ . 
\end{equation}
The other tensor that is of relevance is $t^a_{\alpha\beta}$, whose
only non-vanishing components are 
\begin{equation}\label{tten}
t^0_{\pm\mp} = -\frac12\ , \qquad t^\pm_{\pm\pm} = \pm1\ .
\end{equation}

The triplet algebra is only associative ({\it i.e.}\ satisfies the
Jacobi identity) if certain null states are divided out
\cite{Kausch91,GKau96b}. As a consequence, the algebra only exists at
$c=-2$.  

\subsection{The representation ${\cal R}_1$}

The cyclic states of the indecomposable representation ${\cal R}_1$
are the doublet of states at $h=1$
\be
\phi^\pm = \chi^\pm_{-1} \omega \ .
\ee
These states are not highest weight; instead one has
\begin{eqnarray}
L_1 \, \phi^\pm & = & \chi^\pm_0 \omega = \xi^\pm \nonumber \\
L_{-1} \, \xi^\pm & = & -  \chi^\pm_{-1} \, \Omega  = 
- \psi^\pm   \\
L_0 \, \phi^\pm & = & \phi^\pm + \psi^\pm \ . \nonumber
\end{eqnarray}
To verify this note that the Virasoro modes are
expressed in terms of the fermionic modes as
\be
   L_m ~=~ \sum_{k=-\infty}^{-1} \frac12 \,d_{\alpha\beta}\, 
   \chi^\alpha_k \chi^\beta_{m-k}
   ~+~
   \sum_{k=0}^{\infty} \frac12 \,d_{\alpha\beta} \,
   \chi^\alpha_{m-k} \chi^\beta_{k} ~~.
\labl{eq:Lm-via-chi}
The action of the modes $W^a_n$ can also be easily determined
\cite{GKau96b}. The representation ${\cal R}_1$ is again
indecomposable but reducible; its cyclic vector $\phi^\pm$ is not
highest weight, but the spectrum of $L_0$ is bounded from below
\cite{GabKau96a,GKau96b}, and thus  ${\cal R}_1$ is a highest weight
representation.

\sect{Theta functions}\label{app:theta}

For the description of the free fermionic theory it is convenient to
introduce the functions $f_i$ that are defined as follows 
($q=e^{ 2\pi i \tau}$) 
\begin{eqnarray}
f_1(q) & =& q^{\frac{1}{24}} \prod_{n=1}^{\infty} (1-q^n) \,,
\nonumber\\ 
f_2(q) & = &\sqrt{2}\,
 q^{\frac{1}{24}} \prod_{n=1}^{\infty} (1+q^n) \,,\nonumber \\
f_3(q) & =& q^{-\frac{1}{48}} \prod_{n=1}^{\infty}
\Bigl(1+q^{(n-1/2)}\Bigr) \,,\nonumber\\
f_4(q) & =& q^{-\frac{1}{48}} \prod_{n=1}^{\infty}
\Bigl(1-q^{(n-1/2)}\Bigr) \,. \nonumber
\label{ffunc}
\end{eqnarray}
Under the modular S-transformation, $\tau\mapsto -1/\tau$, 
$f_2$ and $f_4$ get interchanged, while $f_3$ is invariant. The
function $f_1(q)$ agrees with the Dedekind eta-function, and transforms
as  
\be
f_1(-1/\tau) = \sqrt{-i\tau} f_1(\tau) \,.
\ee

\sect{Field insertions and local coordinates}\label{app:local-coord}

To define correlation functions on Riemann surfaces without
a preferred global coordinate one needs a formulation of 
field insertions which includes a local coordinate system
around the insertion point. Below we first introduce the relevant
notation for bulk and boundary fields, and then use this
to compute two-point functions on the Riemann sphere and the
upper half plane.

\subsection{Bulk fields and local coordinates}

To define correlation functions on Riemann surfaces
other than the complex plane, one needs a formulation
of field insertions that does not rely on the existence
of a global coordinate. Let $\Hbulk$ be the space
of bulk states of a conformal field theory, and denote by  
$D_\eps$ the disc 
$\big\{ z\,{\in}\,\Cb\,\big|\,|z|\,{<}\,\eps\big\}$.
A {\em bulk field} on a Riemann surface $\Sigma$ is a pair
\be
  [\varphi,v] \ ; \quad v \in \Hbulk \ , \qquad
  \varphi : D_\eps \rightarrow \Sigma
  ~\text{injective and holomorphic}~.
\labl{eq:bulkfield-def}
In words, $[\varphi,v]$ is
a field labelled by $v$ inserted at the
point $\varphi(0)$ of $\Sigma$ with a choice of
local coordinates $\varphi$ around the insertion point.
A correlator of several bulk fields on a surface
$\Sigma$ is then written as
\be
  \corr{[\varphi_1,v_1] \, [\varphi_2,v_2] 
  \cdots [\varphi_n,v_n]}_\Sigma
  \ . 
\labl{eq:corr-notation}
The ordering of the bulk fields 
can be chosen at will and has no influence on the correlator.
Isomorphic Riemann surfaces
with bulk fields are required to result in equal correlators. 
That is, given an isomorphism $f : \Sigma \rightarrow \Sigma'$,
\be
  \corr{[\varphi_1,v_1] \cdots [\varphi_n,v_n]}_\Sigma
  =
  \corr{[f \circ \varphi_1,v_1] \cdots 
  [f \circ \varphi_n,v_n]}_{\Sigma'}
  \ . 
\labl{eq:corr-covariant}
This holds in particular if $\Sigma=\Sigma'$.

The description \erf{eq:bulkfield-def} of bulk fields is
in fact quite redundant because a change in the local
coordinates can be traded for a change of the vector $v$
without affecting the value of a correlator. This can be
captured by introducing an equivalence relation on bulk fields,
namely we set, for any injective holomorphic function 
$f : D_\eps \rightarrow D_\eps$ with $f(0)=0$,
\be
  [\varphi \circ f , v]
  \sim [\varphi , \rho(f)v ] \ , \qquad
  \text{where} \quad
  \rho(f) = (a_0)^{L_0} (a_0^*)^{\Lo_0} 
  e^{\sum_{m=1}^\infty(a_m L_m + a_m^* \Lo_m)} \ .
\labl{eq:AutO-act}
The $f$-dependent constants $a_m$ are determined by
matching coefficients of
\be
  a_0 \exp\Big(\sum_{m=1}^\infty a_m t^{m+1} 
  \frac{\partial}{\partial t}\Big) t = \sum_{k=0}^\infty
  \frac{1}{k!}\, f^{(k)}(0) \, t^k \ , \qquad
  f^{(k)}(t) = \frac{\partial^k}{\partial t^k} f(t) 
\labl{eq:coord-coeff}
order by order in $t$. The first three coefficients are
\be
  a_0 = f' \ , \quad 
  a_1 = \frac{1}{2}\,\frac{f''}{f'} \ , \quad
  a_2 = \frac{1}{6}\,\frac{f'''}{f'} - 
        \frac{1}{4}\,\Big(\frac{f''}{f'}\Big)^2 \ ,
\ee
with all derivatives evaluated at zero. Replacing a field
$[\varphi,v]$ by an equivalent field $[\psi,w] \sim [\varphi,v]$
does not affect the value of a correlator.

There is no closed expression for the coefficients $a_m$
in \erf{eq:coord-coeff}, but they
can be computed by a recursive formula \cite{Gaberdiel:1994fs}.
For more details on the action of the group of local
coordinate changes and the relation to the definition
of conformal blocks, see section 6.3.1 of 
\cite{Frenkel-BenZvi-2nd-Edition}. 

For \erf{eq:AutO-act} to be well-defined,
the conformal field theory has to fulfil two further
requirements. First,  for every element $v \in \Hbulk$  
there has to exists an $M$ such that $v$ is annihilated by
all $L_m$, $\Lo_m$ with $m>M$. Otherwise $\rho(f)u$ can
result in an infinite sum and is no longer an element of 
$\Hbulk$. Second, consider the family of local coordinates 
$f_t(\zeta) = e^{it} \zeta$. One has $\rho(f_t) = e^{it(L_0-\Lo_0)}$ and
continuously varying $t$ from $0$ to $2\pi$ in
$[\varphi \circ f_t,u] \sim [\varphi,\rho(f_t)u]$ results
in $[\varphi,u] \sim [\varphi,e^{2\pi i (L_0-\Lo_0)}u]$,
which can only be true inside every correlator if
$e^{2\pi i (L_0-\Lo_0)}u = u$. This has to hold for all 
$u\in \Hbulk$.

We can define mode integrals of a holomorphic
field $K \in \Hbulk$ of weight $h$ as
\be
  \corr{\cdots [\varphi, K_m v]
  \cdots}
  = \oint w^{m+h-1}
\corr{\cdots [\varphi \circ (\zeta \mapsto \zeta{+}w) , K]
  \,[\varphi , v]
  \cdots} \frac{dw}{2 \pi i} \ , 
\ee
where the contour integral is along a small circle around
zero that lies entirely in $D_\eps$. By analytic continuation,
the contour integral can then be deformed on the entire Riemann
surface $\Sigma$. Contour integrals of anti-holomorphic
fields are defined similarly.

\subsection{Boundary fields and local coordinates}

For boundary fields the discussion is analogous to the one for bulk
fields. Let $\Hc^\text{bnd}$ be the space
of states of a boundary conformal field theory, and denote 
by $H_\eps$ the half-disc   
$\big\{ z\,{\in}\,D_\eps\,\big|\,\text{Im} z\,{\ge}\,0\big\}$.
A {\em boundary field} on a Riemann surface $\Sigma$ is a pair
$[\varphi,v]$ with $v \in \Hc^\text{bnd}$ and 
$\varphi : H_\eps \rightarrow \Sigma$ an injective, holomorphic and
boundary preserving map. By boundary preserving we mean that
$\varphi$ maps the points $H_\eps \cap \Rb$ to $\partial\Sigma$.

For correlators the notation \erf{eq:corr-notation} is used, and the
identity \erf{eq:corr-covariant} applies also in the presence of
boundary fields. As for bulk fields, the description of boundary
fields is redundant, and a change in the local
coordinates can be traded for a change of the vector $v$
without affecting the value of a correlator. The corresponding
equivalence relation is, for any injective holomorphic function 
$f : H_\eps \rightarrow H_\eps$ with $f(0)=0$,
\be
  [\varphi \circ f , v]
  \sim [\varphi , \rho(f)v ] \ , \quad
  \text{where} \qquad
  \rho(f) = (a_0)^{L_0}  
  e^{\sum_{m=1}^\infty a_m L_m} \ .
\labl{eq:AutO-act-bnd}
The coefficients $a_m$ are the same as in \erf{eq:coord-coeff}.

\subsection{Two-point function on the Riemann sphere}\label{app:2pt-C}

Consider the Riemann sphere $\Pb^1$ and let 
$[z:w]$ be the homogeneous coordinates. Define
the local coordinates $\varphi_a(\zeta) = [\zeta+a:1]$ for
$a \in \Cb$, 
and $\varphi^\infty(\zeta) = [1:\zeta]$. Using the
correlator of two bulk fields on the sphere, we define
a bilinear form $B : \Hbulk \times \Hbulk \rightarrow \Cb$ as
\be
  B(u,v) = \corr{
  [\varphi^\infty,u]\,[\varphi_0,v]}_{\Pb^1} \ .
\labl{eq:B-def}
Composing with the isomorphism $f([z:w]) = [w:z]$ from
$\Pb^1$ to itself and making use of 
the property \erf{eq:corr-covariant} we find
\be
  B(u,v) 
  = \corr{
  [f \circ \varphi^\infty,u]\,[f\circ \varphi_0,v]}_{\Pb^1}
  = \corr{
  [\varphi_0,u]\,[\varphi^\infty,v]}_{\Pb^1} = B(v,u) \ ,
\ee
{\it i.e.}\ $B$ is symmetric. Furthermore, one requires 
$B$ to be non-degenerate, as fields $u$ for which
$B(u,\cdot)$ is identically zero would vanish inside
all correlators and one could pass to a quotient of $\Hbulk$.

In terms of the contour integration mentioned in the
previous section, it is not difficult to verify
that for Virasoro primary holomorphic and anti-holomorphic
fields $K$ and $\ol K$ of weights $h$ and $\overline h$ we have
\be
  B(K_m u, v) = (-1)^h B(u,K_{-m} v) \quad \text{and} \quad
  B(\overline K_m u, v) = (-1)^{\overline h} 
  B(u,\overline K_{-m} v) \ .
\ee
For the symplectic fermion theory, 
this identity shows that we can shift pairs of fermion modes 
from one argument to the other,
\be
  B( \chi^\alpha_m \chi^\beta_n u , v)
  = B( u , \chi^{-\beta}_{-n} \chi^{-\alpha}_{-m} v) \ , \qquad 
  B( \bar\chi^\alpha_m \bar\chi^\beta_n u , v)
  = B( u , \bar\chi^{-\beta}_{-n} \bar\chi^{-\alpha}_{-m} v) \ .
\ee

For a correlator on $\Pb^1$ where all fields 
are of the form $[\varphi_z,\phi]$, we will also use the more
conventional notation $\phi(z)$ to denote the fields, 
\be
  \corr{[\varphi_{z_1},\phi_1]  
  \cdots [\varphi_{z_n},\phi_n]}_{\Pb^1} =
  \corr{ \phi_1(z_1) \cdots \phi_n(z_n)}_{\Pb^1}
  \ . 
\ee
In this notation, the bilinear form 
$B(\,\cdot\,,\,\cdot\,)$ determines the two-point function 
$\corr{ \phi(a) \phi'(b) }_{\Pb^1}$ as follows. 
The M\"obius transformation $M([z:w]) = [z-bw: z-aw]$ 
takes $[a:1]$ to $[1:0]$ and $[b:1]$ to $[0:1]$. Note that
\be
  M \circ \varphi_a(\zeta) 
  = [\zeta{+}a{-}b : \zeta] 
  = \varphi^\infty \circ f(\zeta) \ , \quad
  \text{where} \qquad
  f(\zeta) = \frac{\zeta}{\zeta+a-b}  
\labl{eq:2pt-calc}
and similarly 
$M \circ \varphi_b = \varphi_0 \circ g$
with $g(\zeta) = \tfrac{\zeta}{\zeta+b-a}$.
Using relation \erf{eq:corr-covariant} one computes
\bea
  \corr{ \phi(a) \phi'(b) }_{\Pb^1}
  = \corr{
  [M \circ \varphi_a,\phi]\,[M \circ \varphi_b,\phi']}_{\Pb^1}
  \enl
  \qquad = \corr{
  [\varphi^\infty \circ f,\phi]\,[\varphi_0 \circ g,\phi']}_{\Pb^1}
  = B\big(\rho(f)\phi , \rho(g)\phi'\big) \ .
\eear\labl{eq:omega-two-point}

Let us evaluate this expression for states 
$\bomega$ and $\bOmega$ of the space of bulk states $\Hbulk$ of the
triplet theory. Since $L_0 \bomega = \bOmega = \Lo_0 \bomega$
and $L_0 \bOmega = 0 = \Lo_0 \bOmega$
we have
\be
  B(\bOmega,\bOmega) = B(L_0 \bomega,\bOmega) 
= B(\bomega,L_0 \bOmega) = 0 \ .
\ee
By similar reasoning one sees that $B(\bOmega,\cdot)$ can
only be non-vanishing on the subspace $\Cb \bOmega \oplus \Cb \bomega$
of $\Hbulk$. Non-degeneracy of $B$ then requires
$B(\bOmega,\bomega) \neq 0$. If also $B(\bomega,\bomega) \neq 0$ we
redefine $\bomega \rightarrow \bomega - 
\tfrac{B(\omega,\omega)}{2B(\Omega,\omega)} \bOmega$ so that
$B(\bomega,\bomega) = 0$. 
We take the correlators on $\Pb^1$ to be normalised such that
$B(\bomega,\bOmega) = -1$, so that altogether we get
\be
  B(\bOmega,\bOmega) = B(\bomega,\bomega) = 0
  \ , \qquad B(\bomega,\bOmega) = -1 \ .
\ee
The next ingredient we need in
\erf{eq:omega-two-point} is $\rho(f)u$. Combining
\erf{eq:AutO-act} and \erf{eq:2pt-calc} with 
$L_0 \bomega = \Lo_0 \bomega$ one finds
\be
 \rho(f) \bOmega = \bOmega
 \ , \quad
 \rho(f) \bomega = e^{ \log(|f'|^2) L_0 } \bomega
 = \bomega + \log(|a{-}b|^{-2}) \bOmega \ . 
\ee
For $\rho(g)$ one finds the same result, and 
the two-point functions on $\Pb^1$ are thus equal to 
$\langle \bOmega(a) \bOmega(b) \rangle_{\Pb^1} =
\langle \one \rangle_{\Pb^1} = 0$ and
\be
  \corr{\bomega(a)\bomega(b)}_{\Pb^1}
  = 4 \log|a{-}b| \ , \qquad 
  \corr{\bomega(a)\bOmega(b)}_{\Pb^1} 
  = \corr{\bomega(a)}_{\Pb^1} = -1 \ ,
\labl{eq:omom-bulk-2pt}
where we used that $\bOmega$ is the identity field,
and $\langle \one \rangle$ denotes a correlator with no field
insertions. We are still free to choose the normalisation of the bulk
field $\bmu$. We demand that $B(\bmu,\bmu)=1$ which results in
the two-point function
\be
  \corr{\bmu(a)\bmu(b)}_{\Pb^1}
  = |a{-}b|^{\frac{1}{2}} \ .
\ee
If we combine this with the one-point function of $\bomega$
in \erf{eq:omom-bulk-2pt}, we see that equivalently we can
demand the OPE of $\bmu$ with itself to be of the form
$\bmu(z) \bmu(0) = |z|^{\frac12}\big( - \bomega(0) + 
(\text{other fields})\big)$, which is indeed the convention used
in \erf{eq:bulk-OPE-result}.

\subsection{Two-point function on the upper half
plane}\label{app:2pt-uhp} 

Let $\Ub$ be the upper half plane together with the point
at infinity. We will represent $\Ub$ as the quotient of $\Pb^1$
by the anti-holomorphic involution $\iota : [z:w] \mapsto [z^*:w^*]$.
For points of $\Ub$ we also use the notation $[z:w]$.
The fixed points of $\iota$, which form the boundary of $\Ub$,
are $[r:1]$ for $r \in \Rb$ together with $[1:0]$. We fix an orientation
on $\Ub$ by demanding the map $z \mapsto [z:1]$ from the upper half
plane to $\Ub$ to be orientation preserving.

Let $\tilde\varphi_a : H_\eps \rightarrow \Ub$, $a\in\Rb$, 
and $\tilde\varphi^\infty : H_\eps \rightarrow \Ub$ 
be the local coordinates
$\tilde\varphi_a(\zeta) = [\zeta+a:1]$ and 
$\tilde\varphi^\infty(\zeta) = [1:-\zeta]$. 
One checks that $\tilde\varphi_a$ and $\tilde\varphi^\infty$ are
orientation and boundary preserving (while $\zeta \mapsto [1:\zeta]$
would not be orientation preserving as a map from $H_\eps$ to $\Ub$).

Similar as for bulk fields,
the two-point function on $\Ub$ gives rise to a bilinear pairing $b$
on the space of boundary fields $\Hc^\text{bnd}$
for a given boundary condition,
\be
  b(u,v) = \corr{
  [\tilde\varphi^\infty,u]\,[\tilde\varphi_0,v]}_{\Ub} \ .
\labl{eq:b-def}
Applying the orientation preserving isomorphism $[z:w] \mapsto [w:-z]$
of $\Ub$ shows that $b$ is symmetric, $b(u,v) = b(v,u)$.

For the triplet theory, consider a boundary condition such that
$\Hc^\text{bnd} = \Rc_0$. We then always define the boundary
field $\omega$ such that $L_0 \omega = \Omega$ and such that
$b(\omega,\omega) = 0$. That $b(\Omega,\Omega)=0$ follows in the
same way as it did for the corresponding bulk fields. Using
appropriate M\"obius transformations and \erf{eq:AutO-act-bnd},
the relation between the two-point function of $\omega$ and the
constant $\langle \omega(0) \rangle_{\Ub} = b(\Omega,\omega)$ 
one finds is, for $x,y\in\Rb$, $x>y$,
\be
  \corr{ \omega(x) \omega(y) }_{\Ub} \equiv
  \corr{ [\tilde\varphi_x,\omega]\,[\tilde\varphi_y,\omega] }_{\Ub}
  = - 2 \log(x{-}y)\, \langle \omega(0) \rangle_{\Ub} \ .
\ee

\sect{Intertwiners for fermion representations}

For the purposes of this paper we will need two of the 
indecomposable representations 
of the symplectic fermion mode algebra, namely $\Hc_\om$ and $\Hc_\mu$.
Recall from section \ref{summary} that $\Hc_\om$ is an untwisted
representation 
generated by a vector $\om$ and characterised by the properties that
$\chi^\alpha_m \om = 0$ for $m \in \Zb_{>0}$ and that the four vectors
$\om$, $\chi^+_0 \om$, $\chi^-_0 \om$, $\chi^+_0\chi^-_0 \om$
are linearly independent. The representation $\Hc_\mu$ is twisted
and generated by a vector $\mu$ which obeys $\chi^\alpha_m \mu = 0$ 
for $m \in \tfrac12+\Zb_{\ge 0}$.

Denote by $\Hc^0_\om$ the subspace of $\Hc_\om$ spanned by
$\{ \om, \chi^+_0 \om, \chi^-_0 \om, \chi^+_0\chi^-_0 \om\}$, 
{\it i.e.}\ the
space of lowest generalised $L_0$-weight. Let further 
$Q : \Hc_\om \rightarrow \Hc^0_\om$ be the projector onto $\Hc^0_\om$.
The parity operator on $\Hc_\om$ and $\Hc_\mu$ is denoted by
$(-1)^F$. It is defined by $(-1)^F \chi_m^\alpha = - \chi_m^\alpha (-1)^F$
and $(-1)^F \om = \om$, $(-1)^F \mu = \mu$.
The representations $\Hc_\om$ and $\Hc_\mu$ are graded by generalised
$L_0$ weight, and we will write $\Hc_\om^\vee$ and $\Hc_\mu^\vee$ for
their graded duals.

In the following we shall consider three
types of intertwiners between these representations. 

\subsection{Intertwiners of type 
$\Hc_\om \times \Hc_\om \rightarrow \Hc_\om$}\label{app:ooo-inter}

The (super) vertex algebra built from the symplectic fermions (as well
as its bosonic subalgebra, the triplet vertex operator algebra) are
described in \cite{Abe}. Since the action of $L_0$ on $\Hc_\om$ is not
diagonalisable, an 
intertwiner of type $\Hc_\om \times \Hc_\om \rightarrow \Hc_\om$ can
be logarithmic. Logarithmic intertwiners of representations of vertex
algebras have been treated in 
\cite{Milas:2001bb,Huang:2003za}.
An intertwiner $V$ of type
$\Hc_\om \times \Hc_\om \rightarrow \Hc_\om$ is a linear map 
\be
  V(\,\cdot\,,z) : \Hc_\om \longrightarrow 
  L(\Hc_\om,\Hc_\om)[\![z^{\pm 1},\log z]\!]
\ee
which is compatible with the action of the fermion modes, see 
\cite{Milas:2001bb,Huang:2003za} for details.
Here $L(U,V)$ denotes the space of linear maps between two vector spaces
$U$ and $V$, and by $W[\![x,y,\dots]\!]$ we denote the space of
formal power series in $x,y,\dots$ with coefficients in $W$. 
To be completely precise, we should write $\ell_z$ instead of $\log z$
and treat it as an independent formal variable with certain properties,
but this distinction will not be important below.

The statement that $V(\,\cdot\,,z)$ is an intertwiner is equivalent to
the identities, for $u \in \Hc_\om$, $a \in \Hc^0_\om$ and $m \in \Zb$,
\be\begin{array}{rl}\displaystyle
V\big(\chi^\alpha_m u ,z\big) \etb= \sum_{k=0}^\infty \binom{m}{k}
\Big( (-z)^k \chi^\alpha_{m-k} V\big(u,z\big)
- (-z)^{m-k} V\big((-1)^F u,z\big) \chi^\alpha_{k} \Big) \ ,
\enl
\chi_m^\alpha V\big(a ,z\big) \etb= 
z^m V\big(\chi^\alpha_0 a ,z\big)
+ V\big((-1)^F a ,z\big) \chi_m^\alpha \ ,
\enl
\big[L_{-1},V\big(u ,z\big)\big] \etb= 
\frac{\partial}{\partial z} V\big(u ,z\big) \ .
\eear\labl{eq:V-modecom}
These commutation relation for the modes can be derived from the usual 
contour deformation arguments. A useful relation which is a direct
consequence of \erf{eq:V-modecom} 
is
\be
  [\chi^\alpha_m,V(\om,z)] = z^m [\chi^\alpha_0,V(\om,z)] \ .
\ee
For $u^* \in \Hc_\om^\vee$ and $v \in \Hc_\om$ denote by
$\langle u^*,v\rangle$ the canonical pairing.
Using the relations \erf{eq:V-modecom}, the expression
$\langle u^*, V(v,z)w \rangle$, for $u^* \in \Hc_\om^\vee$, 
$v,w \in \Hc_\om$, can be reduced to a finite sum involving 
only terms of the form
$\langle a^*, V(\om,z)b \rangle\, z^m (\log z)^n$, where
$a^* \in (\Hc^0_\om)^*$, $b \in \Hc^0_\om$ and $m,n \in \Zb$.

Abbreviate by $V^0(u,z) = Q V(u,z) Q$ the restriction of $V(u,z)$ to
$\Hc^0_\om$. Then the above reasoning implies that $V(u,z)$ is 
uniquely determined by $V^0(\om,z)$ and the relations \erf{eq:V-modecom}.
By analysing the action of $L_0$ or by directly using 
\cite[proposition 3.9]{Huang:2003za} we see,
\be
  V^0(\om,z) = A + B \log z + C (\log z)^2 + D (\log z)^3 \ ,
\ee
where $A,B,C,D \in L(\Hc^0_\om,\Hc^0_\om)$. Inserting the expression
\erf{eq:Lm-via-chi} of $L_{-1}$ 
in terms of fermion modes into the relation
$Q [L_{-1},V^0(\om,z)] Q = \tfrac{\partial}{\partial z} V^0(\om,z)$
gives the condition
\bea
  z^{-1}\Big( \chi^-_0 V^0(\om,z) \chi^+_0 
     - \chi^+_0 V^0(\om,z) \chi^-_0 + 
         2 V^0(\om,z) \chi^+_0 \chi^-_0 \Big)
  \enl    \quad      
  = z^{-1} B + 2 C z^{-1} \log z + 3 D z^{-1} (\log z)^2 \ .
\eear\ee
This allows to fix $B,C,D$ in terms of $A$. 
Thus giving an element $A \in L(\Hc^0_\om,\Hc^0_\om)$ determines uniquely
an intertwiner of type $\Hc_\om \times \Hc_\om \rightarrow \Hc_\om$.
We will denote this intertwiner by $V_A(\,\cdot\,,z)$. Its action on
$\Hc^0_\om$ is easily found to be 
(recall that $\Om = -\chi^+_0 \chi^-_0 \om$)
\be\begin{array}{rl}\displaystyle
  V^0_A(\om,z) \etb= A + \big(\chi^-_0 A \chi^+_0 - \chi^+_0 A \chi^-_0 + 
2 A \chi^+_0 \chi^-_0 \big) \log z 
- \chi^+_0 \chi^-_0 A \chi^+_0 \chi^-_0 (\log z)^2 \ ,
\enl
  V^0_A(\chi^\alpha_0\om,z) \etb= \chi^\alpha_0 A - A \chi^\alpha_0 +
    \big( \chi^+_0\chi^-_0 A \chi^\alpha_0 + 
    \chi^\alpha_0 A  \chi^+_0\chi^-_0 \big) \log z \ ,
  \enl
  V^0_A(\Om,z) \etb= -\chi^+_0\chi^-_0 A + \chi^+_0 A \chi^-_0 
     - \chi^-_0A \chi^+_0 - A \chi^+_0\chi^-_0  \ .
\eear\labl{eq:V0A-args}
Denote by $\text{Hom}(\Hc_\om \otimes \Hc_\om, \Hc_\om)$ 
the space of intertwiners of type 
$\Hc_\om \times \Hc_\om \rightarrow \Hc_\om$. The above arguments show
that this space is at most 16 dimensional (which is the dimension of 
$L(\Hc^0_\om,\Hc^0_\om)$). Since $\Hc_\om$ does not contain null-vectors
(with respect to the fermion modes),
it is plausible that the dimension is exactly 16, but we have not pursued
this further. 

\subsection{Intertwiners of type 
$\Hc_\mu \times \Hc_\mu \rightarrow \Hc_\om$
and $\Hc_\mu \times \Hc_\om \rightarrow \Hc_\mu$}

Let $U$ be an intertwiner of 
type $\Hc_\mu \times \Hc_\mu \rightarrow \Hc_\om$,
\be
  U(\,\cdot\,,z) : \Hc_\mu \longrightarrow 
  z^{\frac14}L(\Hc_\mu,\Hc_\om)[\![z^{\pm 1},\log z]\!] \ .
\ee
The condition analogous to \erf{eq:V-modecom} is more involved for
twisted representations, see {\it e.g.}\ 
\cite{KW,DLM,Gaberdiel:1996kf}. We will just need the relation
\be
  \chi^\alpha_m U(\mu,z)\mu 
  = - \sum_{k=1}^\infty \binom{\frac12}{k}(-z)^k 
            \chi^\alpha_{m-k} U(\mu,z)\mu
  = \Big( \tfrac{z}{2} \chi^\alpha_{m-1} 
       + \tfrac{z^2}{8} \chi^\alpha_{m-2} + \cdots
                       \Big)  U(\mu,z)\mu \ ,
\labl{eq:U-modecom}
which is valid for $m \ge 1$. It
can be derived by expanding
\be
  0 = \oint_{\mathcal{C}_\infty} \frac{d\zeta}{2 \pi i}
  \sqrt{\zeta(\zeta{-}z)} \zeta^{m-1} \chi^\alpha(\zeta) 
      \, U(\mu,z)\, \mu 
\labl{eq:U-modecom-aux1}
in powers of $\zeta^{-1}$. The contour
$\mathcal{C}_\infty$ is a large anti-clockwise oriented circle containing
the points $0$ and $z$.
In any case, the expression $\langle u^*, U(v,z) w \rangle$ for
$u^* \in \Hc_\om^\vee$ and $v,w \in \Hc_\mu$ is uniquely determined once
we know $Q U(\mu,z)\mu$. Compatibility with the action of $L_0$ forces
\be
  Q U(\mu,z)\mu = a + b \log z \ ,
\ee
for $a,b \in \Hc^0_\om$. As before, $b$ is fixed in terms of $a$ by
exploiting the condition 
$[L_{-1},U(\mu,z)] = \tfrac{\partial}{\partial z}U(\mu,z)$.
This is easiest done in the form
$[2 z^{-1} L_0 - z^{-2}L_1,U(\mu,z)] = 
\tfrac{\partial}{\partial z}U(\mu,z)$,
together with \erf{eq:U-modecom}. 
Denoting the intertwiner determined by 
$a \in \Hc^0_\om$ by $U_a(\,\cdot\,,z)$ we find
\be
  Q U_a(\mu,z) \mu = z^{\frac14}\big(a 
         - \log z \cdot \chi^+_0 \chi^-_0 a\big) \ .
\labl{eq:U-leading}
It also follows that the dimension of 
$\text{Hom}(\Hc_\mu \otimes \Hc_\mu, \Hc_\om)$ is at most four.
\smallskip

\noindent Next, let $\Ut$ be an intertwiner of 
type $\Hc_\mu \times \Hc_\om \rightarrow \Hc_\mu$,
\be
  \Ut(\,\cdot\,,z) : \Hc_\mu \longrightarrow 
  L(\Hc_\om,\Hc_\mu)[\![z^{\pm 1},\log z]\!] \ .
\ee
Let $\mu^* \in \Hc_\mu^\vee$ be the unique linear form such that
$\langle\mu^*,\mu\rangle = 1$ and $\langle\mu^*,v\rangle = 0$ for
any $v \in \Hc_\mu$ with $L_0$-weight greater than $-\tfrac18$.
The relation corresponding to \erf{eq:U-modecom} reads
\be
  \mu^* \circ \Ut(\mu,z) \chi^\alpha_m 
  = - \sum_{k=1}^\infty \binom{\frac12}{k}(-z)^{-k}
    \mu^* \circ \Ut(\mu,z)\chi^\alpha_{m+k} \ ,
\labl{eq:Ut-modecom}
which is valid for $m \le -1$.
The expression $\langle u^*, \Ut(v,z) p \rangle$ for
$u^* \in \Hc_\mu^\vee$, $v \in \Hc_\mu$ and $p \in \Hc_\om$ is 
then uniquely determined once
we know $\mu^* \circ \Ut(\mu,z) Q$. 
Compatibility with the action of $L_0$ and $L_{-1}$ shows that 
an intertwiner of type $\Hc_\mu \times \Hc_\mu \rightarrow \Hc_\om$ is 
determined by an element $\varphi \in (\Hc^0_\om)^*$. The corresponding
intertwiner is denoted by $\Ut_\varphi(\,\cdot\,,z)$ and obeys 
\be
  \mu^* \circ \Ut_\varphi(\mu,z) Q = \varphi + \log z \cdot
  \varphi \chi^+_0 \chi^-_0 \ .
\labl{eq:Ut-leading}
The dimension of 
$\text{Hom}(\Hc_\mu \otimes \Hc_\om, \Hc_\mu)$ is therefore also at
most four.

\sect{Fermionic four-point blocks}\label{app:four-point-blocks}

For the computations in the main text we need various bosonic
four-point blocks, which all derive from three basic fermionic
four-point blocks, namely the blocks 
involving four times $\Hc_\om$, two times 
$\Hc_\om$ and two times $\Hc_\mu$, and
four times $\Hc_\mu$.

\subsection{The $\om\om\om\om$-block}\label{app:oooo-block}

We need to compute
\be
  Q \, V_A(a,z) \, (-1)^{\eps F} \, V_B(b,w) \, Q 
  \  \in \ L(\Hc^0_\om,\Hc^0_\om)
  [\![z^{\pm 1},w^{\pm 1},\log z,\log w]\!] \ ,
\labl{eq:oooo-aux1}
where $A,B \in L(\Hc^0_\om,\Hc^0_\om)$, $a,b \in \Hc^0_\om$ and
$\eps \in \{0,1\}$. We will do this by explicitly summing over all 
intermediate states. This is possible because
\be
  \chi^\alpha_k \chi^\beta_l \chi^\gamma_m V_B(b,w) Q = 0
\ee
for all $m,n,k \ge 1$, as follows by applying \erf{eq:V-modecom} three
times and noting that a product of three fermion zero modes is always zero.

One can verify that the identity map on $\Hc_\om$ can be written as
\be
  \id_{\Hc_\om} 
  = Q - \sum_{\alpha = \pm} \sum_{m=1}^\infty \frac{\alpha}{m}
    \chi^\alpha_{-m} Q \chi^{-\alpha}_{m}
    - \sum_{m,n=1}^\infty  \frac{1}{m\,n} 
    \chi^+_{-m} \chi^-_{-n} Q \chi^+_{n} \chi^-_{m}
    + R \ ,
\labl{eq:H_om-id-expand}
where $R$ contains only terms with three or more fermion modes on both
sides of the projector $Q$. Inserting this into \erf{eq:oooo-aux1} gives
\bea
  Q \, V_A(a,z) \, (-1)^{\eps F} \, V_B(b,w) \, Q
  \enl
  \quad =~ V^0_A(a,z) (-1)^{\eps F} V^0_B(b,w)
  \enl \qquad \qquad 
    - \sum_{\alpha = \pm} \sum_{m=1}^\infty \frac{\alpha}{m}
    \Big(\frac wz\Big)^m (-1)^\eps  V^0_A((-1)^F \chi^\alpha_0 a,z) 
    (-1)^{\eps F} V^0_B(\chi^{-\alpha}_0 b,w)
  \enl \qquad \qquad 
    - \sum_{m,n=1}^\infty  \frac{1}{m\,n} \Big(\frac wz\Big)^{m+n}
    V^0_A(\chi^+_0 \chi^-_0 a,z) (-1)^{\eps F} 
                  V^0_B(\chi^+_0 \chi^-_0 b,w)
  \enl
  \quad=~ V^0_A(a,z) (-1)^{\eps F} V^0_B(b,w)
  \enl \qquad \qquad 
    + (-1)^\eps \log\frac{z{-}w}z \sum_{\alpha = \pm}
    \alpha \, V^0_A((-1)^F \chi^\alpha_0 a,z) 
    (-1)^{\eps F} V^0_B(\chi^{-\alpha}_0 b,w)
  \enl \qquad \qquad 
    -\Big( \log\frac{z{-}w}z \Big)^2
    V^0_A(\chi^+_0 \chi^-_0 a,z) 
         (-1)^{\eps F} V^0_B(\chi^+_0 \chi^-_0 b,w) \ .
\eear\labl{eq:oooo-fermblock1}

The same calculation can be carried out to compute the product
of intertwiners in the crossed channel. The result is,
for $a,b \in \Hc^0_\om$ and $C,D \in L(\Hc^0_\om,\Hc^0_\om)$,
\bea
  Q \, V_C\big( \,(-1)^{\eps F} V_D(a,z{-}w) b,\, w\big)\, Q 
  \enl\quad 
  =~ V_C^0\big((-1)^{\eps F} V_D^0(a,z{-}w) b,w\big) 
  \enl \qquad\qquad
  + \log \frac zw \sum_{\alpha=\pm} \alpha V_C^0\big( (-1)^{(\eps+1)(F+1)} 
   V_D^0(\chi^{-\alpha}_0 a,z{-}w)b,w \big) \chi^\alpha_0
  \enl \qquad\qquad
  - \Big( \log \frac zw \Big)^2 ~V_C^0\big((-1)^{\eps F} 
  V_D^0(\chi^+_0 \chi^-_0 a,z{-}w) b,w\big) \chi^+_0 \chi^-_0
     \ .
\eear\labl{eq:oooo-fermblock2}

\subsection{The $\mu\mu\mu\mu$-block}

The next block we are interested in is
\be
  F(x) = \mu^* \circ  \Ut_\varphi(\mu,1) (-1)^{\eps F} U_a(\mu,x) 
           \mu   \ ,
\labl{eq:mmmm-aux1}
where $\varphi \in (\Hc^0_\om)^*$ and $a \in \Hc^0_\om$. 
We have also set $z=1$ for the first intertwiner since this is the
form in which the block is used in the main text.

The block can be computed using the level 2 Virasoro null vector
$N = (L_{-2}  - 2 L_{-1}^2)\mu = 0$ in $\Hc_\mu$. Denote by
$f(x) = \langle \mu| \mu(1) \mu(x) \mu(0) \rangle$ a conformal block
with insertions of $\mu$ at $0,1,x$ and $\infty$. The null vector implies
that
\be
  \langle \mu | N(1) \, \mu(x) \, \mu(0) \rangle
  =0 \ .
\ee
Applying the usual contour deformation arguments 
gives a second order differential equation for $f(x)$, namely
\be
  16 x^2 (x{-}1)^2 f''(x) + 8 x (x{-}1) (2x{-}1) f'(x) + f(x) = 0 \ .
\labl{eq:mmmm-deq}
The space of solutions is spanned by
\be\begin{array}{lll}\displaystyle
  f_1(x) \etb= \big( x(1{-}x) \big)^{\frac14} 
     {}_2 F_1(\tfrac{1}{2},\tfrac{1}{2};1;x)
  \etb= x^{\frac14}\Big( 1 - \tfrac{1}{64} x^2 + \Oc(x^3) \Big)
\enl 
  f_2(x) &= \big( x(1{-}x) \big)^{\frac14} 
     {}_2 F_1(\tfrac{1}{2},\tfrac{1}{2};1;1{-}x)
  \etb= x^{\frac14}\Big( \tfrac{4}{\pi} \log 2 
        - \tfrac{1}{\pi} \log x - 
        \tfrac{1}{2\pi} x + \orl(x^2) \Big) \ ,
\eear\labl{eq:mmmm-aux2}
{\it i.e.}\ we find that
\be
  \langle \mu | \mu(1) \mu(x) \mu(0) \rangle =  C_1 f_1(x) + C_2 f_2(x)
\labl{eq:mmmm-space} 
for some constants $C_1,C_2$.

The product of intertwiners $F(x)$ given in \erf{eq:mmmm-aux1} also
solves the differential equation \erf{eq:mmmm-deq} and therefore has to be
of the form \erf{eq:mmmm-space}. Matching the leading order of $F(x)$ 
(obtained by combining \erf{eq:U-leading} and \erf{eq:Ut-leading})
against the asymptotics of the solutions in \erf{eq:mmmm-aux2} fixes
the constants $C_1$ and $C_2$ in \erf{eq:mmmm-space} uniquely, and one
finds
\bea
  \mu^* \circ  \Ut_\varphi(\mu,1) (-1)^{\eps F} U_a(\mu,x) \mu
  \enl
  \quad =~ \big(x(1{-}x)\big)^{\frac14}
  \Big\{ \varphi\big((-1)^{\eps F}(1- 4 \log 2\cdot 
  \chi^+_0 \chi^-_0)a\big) 
  {}_2 F_1(\tfrac{1}{2},\tfrac{1}{2};1;x)
  \enl \hspace*{8em}
  + \pi  \varphi\big((-1)^{\eps F} \chi^+_0 \chi^-_0 a\big) 
  {}_2 F_1(\tfrac{1}{2},\tfrac{1}{2};1;1{-}x)
  \Big\} \ .
\eear\ee
It follows that the space of blocks of type
$\langle \mu | \mu(1) \mu(x) \mu(0) \rangle$ is given by the product
of intertwiners 
$\mu^* \circ  \Ut_\varphi(\mu,1) (-1)^{\eps F} U_a(\mu,x) \mu$ for
all possible choices of $\varphi$ and $a$. This is in accordance
with the analysis of the 
fusion of representations of the triplet algebra 
\cite{GKau96b}, which implies that only the representation $\Hc_\om$
appears in the fusion of $\Hc_\mu$ with itself.

\subsection{The $\mu\mu\om\om$-block}\label{app:mmoo-block}
Let 
\be
  g(x) = \langle \mu | \mu(1) \om(x) \om(0)\rangle
\labl{eq:mmoo-gdef}   
be a conformal block with states $\om, \om, \mu$ and $\mu$
inserted at $0,1,x$ and $\infty$, respectively. To obtain the differential
equation from inserting the null vector 
$N = (L_{-2}  - 2 L_{-1}^2)\mu$ at the point 1,
the following identity is helpful,
\bea
\langle \mu | \mu(1) \, \psi(x) \, 
(L_{-1}\xi)(0) \rangle = 
\langle \mu | \mu(1) \, \psi(x) \, 
(L_{0}\xi)(0) \rangle + 
\langle \mu | \mu(1) \, (L_0 \psi) (x) \,  \xi(0) \rangle
\enl\hspace*{15em} 
+ (x{-}1)\, 
\langle \mu | \mu(1) \, (L_{-1}\psi)(x) \, \xi(0) \rangle 
\ , 
\eear\ee
where $\psi$ is a Virasoro highest weight state and $\xi$ is arbitrary.
With the help of this identity it is also straightforward to determine
the conformal blocks obtained by replacing some of the $\om$ insertions
in $g(x)$ by $\Om$ insertions,
\bea
  \langle \mu | \mu(1) \Omega(x) \Omega(0) \rangle = C_1
  \ , \qquad
  \langle \mu | \mu(1) \Omega(x) \omega(0) \rangle = C_2 \ ,
  \enl
  \langle \mu | \mu(1) \omega(x) \Omega(0) \rangle 
        = -C_1 \log(1{-}x)+C_3 \ ,
\eear\labl{eq:mmoo-Om-ins}
where $C_1, C_2, C_3 \in \Cb$ are integration constants.
Putting everything together, one computes the following 
differential equation for $g(x)$,
\be
  2 x^2 g''(x) - \frac{x(3x{-}2)}{1{-}x} g'(x) + \frac{4x}{1{-}x} C_1 
  + \frac{x(x{-}2)}{(1{-}x)^2} C_2 = 0 \ .
\labl{eq:mmoo-deq}
It is solved by
\be
  g(x) = - C_1 h(x)^2 - C_2 \log(1{-}x) + C_4 h(x) + C_5  \ , \qquad 
   h(x) = \log \left( \frac{ 1 - \sqrt{1-x}}{1+\sqrt{1-x}} \right) \ . 
\labl{eq:mmoo-gx}
Here $C_4, C_5 \in \Cb$ are two more integration constants.

\medskip

The same considerations which determined the space of function
\erf{eq:mmoo-gx}, to which conformal blocks of the form 
$\langle \mu | \mu(1) \om(x) \om(0)\rangle$ belong, can be used to
determine the product of intertwiners
\be
  G(x) = \mu^* \circ  \Ut_\varphi(\mu,1) (-1)^{\eps F} V_A(a,x) Q
\labl{eq:mmoo-aux1}
with $\varphi \in (\Hc^0_\om)^*$, $A \in L(\Hc^0_\om,\Hc^0_\om)$
and $a \in \Hc^0_\om$. One uses the null vector $N$ as
well as the identity
\be
  \mu^*\mu  V_A(a,x) L_{-1}
  = \mu^*\mu  
       V_A((x{-}1)L_{-1}a + L_0a,x)
    + \mu^*\mu  V_A(a,x) L_0 \ ,
\labl{eq:mmoo-L-1-L0}
where we abbreviated 
$\mu^*\mu \equiv \mu^* \circ  \Ut_\varphi(\mu,1) (-1)^{\eps F}$,
to obtain a differential equation for $G(x)$ analogous to
\erf{eq:mmoo-deq}. The solution one finds is
\be
  G(x) = - h(x)^2 \psi L_0 - \log(1{-}x) \psi + h(x) \xi + \beta  \ , 
  \qquad
  \psi := \varphi \circ (-1)^{\eps F} V^0_A(L_0 a,x) \ ,
\labl{eq:mmoo-Gx}
where $h(x)$ is defined as in \erf{eq:mmoo-gx}, and 
$\psi \in (\Hc^0_\om)^*$ can be checked to be independent of $x$. The
linear forms $\xi,\beta \in (\Hc^0_\om)^*$ can be
fixed by matching the leading term in the small $x$ expansion of
\erf{eq:mmoo-aux1} and \erf{eq:mmoo-Gx}, {\it i.e.}\ by solving
\bea
  \varphi \circ (-1)^{\eps F} V^0_A(a,x)
  \enl
  = \beta - 2 \log 2 \cdot (\xi + 2 \log 2 \cdot \psi L_0) + 
    \log x \cdot (\xi + 4 \log 2 \cdot \psi L_0) 
- (\log x)^2 \psi L_0 \ .
\eear\ee
This determines $\xi$ and $\beta$ uniquely. Setting $a=\omega$,
the product 
$\mu^* \circ  \Ut_\varphi(\mu,1) (-1)^{\eps F} V_A(\omega,x) \omega$ 
solves the differential equation \erf{eq:mmoo-deq}. Moreover, the
constants 
$C_1 = \psi(L_0 \omega)$,
$C_2 = \psi(\omega)$,
$C_4 = \xi(\omega)$ and 
$C_5 = \beta(\omega)$
appearing in \erf{eq:mmoo-gx}
can be chosen independently by appropriately varying $A$ and
$\varphi$, so that all blocks
of type $\langle \mu | \mu(1) \om(x) \om(0)\rangle$ are of the form 
\erf{eq:mmoo-aux1}. This is again in accordance with the fusion
analysis of \cite{GKau96b}, which implies that only $\Hc_\om$ appears
in the fusion of  $\Hc_\mu$ with itself, and in the fusion of
$\Hc_\om$ with itself. 

\medskip

In the crossed channel the four-point block is written in terms of 
intertwiners as
\be
  \tilde G(x) = \mu^* \circ  \Ut_\varphi\big(
  (-1)^{\eps F} \Ut_\psi(\mu,1{-}x)b,x\big) Q
\labl{eq:mmoo-cross1}
for some $\varphi,\psi \in (\Hc^0_\om)^*$ and $b \in \Hc^0_\om$.
One first notes that if $b = L_0 a$ for some $a$, then 
$\tilde G(x)$ does not depend on $x$. We can thus evaluate it in the
limit $x \rightarrow 1$ and find
\be
  \mu^* \circ  \Ut_\varphi\big(
  (-1)^{\eps F} \Ut_\psi(\mu,1{-}x)L_0a,x\big) Q
  = \langle \psi,L_0 a\rangle \varphi \ .
\ee
Again, $\tilde G(x)$ solves a differential equation similar to
\erf{eq:mmoo-deq}, but this time with $C_1$ replaced by
$\langle \psi,L_0 a\rangle \varphi L_0$ and $C_2$ replaced by
$\langle \psi,L_0 a\rangle \varphi$. The solution is
\be
  \tilde G(x) = - h(x)^2 \langle \psi,L_0 a\rangle \varphi L_0 
  - \log(1{-}x) \langle \psi,L_0 a\rangle \varphi + h(x) \xi' 
     + \beta' \ ,
\labl{eq:Gtilde-sol}
where the integration constants $\xi',\beta' \in (\Hc^0_\om)^*$ are fixed
by matching the $x \rightarrow 1$ asymptotics of 
\erf{eq:mmoo-cross1} and \erf{eq:Gtilde-sol}, {\it i.e.}\ by solving
(setting $x = 1{-}\delta$) 
\bea
  \langle \psi,a\rangle \varphi - \log \delta \cdot 
	\langle \psi,L_0 a\rangle \varphi
  + 2\sqrt{\delta} \cdot (-1)^\eps \sum_{\alpha = \pm} 
	\alpha \psi( \chi^{-\alpha}_0 a)
    \varphi \chi^\alpha_0 + \orl(\delta) 
  \enl
  \quad = ~
  \beta' - \log \delta \cdot \langle \psi,L_0 a\rangle \varphi 
  - 2\sqrt{\delta}\, \xi'  + \orl(\delta) \ ,
\eear\ee
from which one can read off $\xi'$ and $\beta'$ directly. To work out
the left hand side one needs to make use of the identities, which hold
for $a\in\Hc^0_\om$, 
\be
  \mu^* \,\chi^\alpha_{\frac12} \, \Ut(\mu,z)a = i \sqrt{z}  \cdot
  \mu^* \,\Ut(\mu,z) \,\chi^\alpha_0 a \ , \quad
  \mu^* \,\Ut(\chi^\alpha_{-\frac12} \mu,z)a = \frac{i}{\sqrt{z}} \cdot
  \mu^* \,\Ut(\mu,z)\, \chi^\alpha_0 a \ ,
\ee
which in turn are obtained by contour deformation arguments.

\subsection{Bosonic intertwiners and blocks}\label{app:boson-blocks}

To obtain the four point blocks for the various representations of the
triplet algebra, one has to project the fermionic representations to
subspaces of fixed fermion number. Let
\be
  P^\eta = \frac12\big( 1 + \eta \, (-1)^F \big) \ .
\labl{eq:project-boson}
Then for example 
\be
  u \mapsto P^+ V_A(u,z) P^+
\labl{eq:om-inter-R0}
with $u$ in $\Rc_0 \subset \Hc_\om$ and $V_A(\,\cdot\,,z)$ as defined
in section \ref{app:ooo-inter} is an intertwiner from 
$\Rc_0 \times \Rc_0$ to $\Rc_0$, {\it i.e.}\ a linear map from
$\Rc_0$ to $L(\Rc_0,\Rc_0)[\![ z^\pm 1, \log z ]\!]$. 

Accordingly, for example the space of four-point blocks with
insertions of $\mu$ at $0,1,x$ and $\infty$ and with $\Rc_1$ running
in the intermediate channel is given by
\be
   \big\langle \mu^* , \Ut_\varphi(\mu,1) P^- U_a(\mu,x) \mu
\big\rangle \ ,
\ee
with $\varphi$ and $a$ taking values in $(\Hc^0_\om)^*$ and
$\Hc^0_\om$, respectively. Note that distinct choices of $\varphi$ and
$a$ do not necessarily result in different 4-point blocks.

\sect{Associativity of the boundary fields} 

\subsection{Associative, unital algebras 
on $\Rc_0$}\label{app:alg-assoc}

The consistency of the OPE of boundary fields on the $(\Nr,\pm)$
boundary with factorisation (or crossing) can also be formulated on
the level of intertwiners. For $u \in \Rc_0$ denote by
\be
  \Lambda_M(u,x) = P^+ V_M(u,x) P^+
\ee
the intertwiner of $\Rc_0$ representations as in equation
\erf{eq:om-inter-R0}, with $P^+$ the projector \erf{eq:project-boson}
on the subspace  $\Rc_0 \subset \Hc_\om$,
and $M \in L(\Hc^0_\om,\Hc^0_\om)$ (see the end of 
section \ref{app:ooo-inter}).
The OPE of two boundary fields 
$\psi, \psi' \in \Hc^\text{bnd} \equiv \Rc_0$ can  
be written as
\be
  \psi(x) \psi'(0) = \Lambda_M(\psi,x) \psi' \ .
\labl{eq:om-ope-inter}
Factorisation of the boundary OPE can now be formulated as an
associativity condition for the intertwiners $V_M$, {\it i.e.}\ one
has to find a $M \in L(\Hc^0_\om,\Hc^0_\om)$ such that
\be
  \Lambda_M(\psi,x) \Lambda_M(\psi',y)
  = \Lambda_M( \Lambda_M(\psi,x{-}y)\psi',y) 
\labl{eq:om-ope-assoc}
for all $\psi,\psi' \in \Rc_0$.
The condition that $\Omega(x)$ should be the identity field in turn reads
\be
  \Lambda_M(\Omega,x) = \id_{\Rc_0} \ .
\labl{eq:om-ope-unit}
Phrased in the language of the representation category of the triplet
algebra, conditions \erf{eq:om-ope-assoc} and \erf{eq:om-ope-unit}
amount to endowing the object $\Rc_0$ with the structure of a unital,
associative algebra. Let us denote this algebra by $A_M$. 
The algebra $A_M$ is a logarithmic analog of the special symmetric
Frobenius algebras used in \cite{Fuchs:2002cm} to describe local
conformal field theories, or of the open string vertex algebras of
\cite{Huang:2003ju}. 

The solutions to \erf{eq:om-ope-assoc} and \erf{eq:om-ope-unit} can be
written out explicitly. To find them one uses the conformal blocks
given in \erf{eq:oooo-fermblock1} and \erf{eq:oooo-fermblock2}. This
results in a number of linear and quadratic relations for $M$ (which
one can conveniently keep track of with a computer algebra
program). Let us choose the basis 
\be
 v_1 = \omega \ ,\qquad
 v_2 = \chi^+_0 \omega \ ,\qquad
 v_3 = \chi^-_0 \omega \ , \qquad
 v_4 = \chi^+_0 \chi^-_0 \omega
\labl{eq:H0om-basis}
of $\Hc_\om$, so that we can express $M$ as a $4{\times}4$-matrix
$(m_{ij})$. Because of the projectors, the intertwiner 
$\Lambda_M(\,\cdot\,,z)$ will only depend on the 8 entries
\be
  M = \begin{pmatrix}  
    m_{11} & * & * & m_{14} \\  
    *   & m_{22} & m_{23} & * \\  
    *   & m_{32} & m_{33} & * \\  
    m_{41} & * & * & m_{44} 
 \end{pmatrix}
\ee
of $M$. Requiring 
\erf{eq:om-ope-assoc} and \erf{eq:om-ope-unit} to hold is now equivalent
to (upon setting the irrelevant entries of $M$ to zero)
\be
  M = \begin{pmatrix}  
    m_{11} & 0 & 0 & -1 \\  
    0   & m_{22} & m_{23} & 0 \\  
    0   & m_{32} & m_{11}{-}m_{22} & 0 \\  
    m_{22}(m_{11}{-}m_{22}) - m_{23}m_{32} & 0 & 0 & 0 
 \end{pmatrix} \ .
\labl{eq:M-general}
In particular, there are four free parameters 
$m_{11}, m_{22}, m_{23}, m_{32} \in \Cb$. Some of these solutions will
be isomorphic in the sense that they can be related via 
\be
  \Lambda_{M'}(u,z) = f^{-1} \Lambda_M(f u,z) f \ ,
\labl{eq:M-redefine}
where $f : \Rc_0 \rightarrow \Rc_0$ is an isomorphism of
representations. Since $f$ is determined uniquely by $f(\om)$, the
space of such isomorphisms is two-dimensional. In addition, $f$ has
to preserve the unit of the algebra, $f(\Om)=\Om$, which leaves a
one-dimensional space. The freedom \erf{eq:M-redefine} can thus be
used to remove one of the parameters, for example one can always
achieve $m_{11}=0$. The leading term in the OPE of $\omega$ with
itself then reads 
\be
  \om(x) \om = \Lambda_M(\omega,x) \omega = - 2 \log x \cdot \om + 
  \big((m_{22})^2{+}m_{23} m_{32} - (\log x)^2\big) \, \Om + 
  \orl[h{=}1,x]\ ,
\labl{eq:omom-ope-via-M}
which is indeed of the form \erf{eq:omom-bnd-ope-aux1}. 

\medskip

For non-logarithmic rational conformal field theories one finds
\cite{Fuchs:2002cm,Fjelstad:2005ua} that {\em every} algebra of
boundary fields with certain extra properties (it has to be special
symmetric Frobenius) gives rise to a consistent local conformal field
theory.  The construction of \cite{Fuchs:2002cm,Fjelstad:2005ua} 
does not apply in the present case, but it would nonetheless be
interesting to investigate if also other solutions of the family 
\erf{eq:M-general} describe the algebra of boundary fields
for a consistent conformal field theory (for which the chiral symmetry
contains the triplet algebra). 

\subsection{$\Vc_{-\frac18}$ as a representation of the boundary
algebra}\label{app:V18-rep} 

The analysis of the cylinder partition functions in section
\ref{sec:cardy-sol} showed that the space boundary changing fields
form $(\Dr,\eta)$ to $(\Nr,\eta)$ is isomorphic to
$\Vc_{-\frac18}$. For  $u \in \Vc_{-\frac18}$ and 
$\rho \in (\Hc^0_\om)^*$ denote by 
\be
  R_\rho(u,z) = P^+ \Ut_\rho(u,z) P^+ 
\ee
the intertwiner of type 
$\Vc_{-\frac18} \times \Rc_0 \rightarrow \Vc_{-\frac18}$ 
obtained by projecting $\Ut_\rho(\,\cdot\,,z)$.
The OPE of a boundary changing field $\xi(x)$
with a boundary field $\psi(0)$ on the $(\Nr,\eta)$ boundary can then be
written as 
\be
  \xi(x) \psi(0) = R_\rho(\xi,x) \psi \ ,
\ee
and the factorisation condition takes the form
\be
  R_\rho(R_\rho(\xi,x{-}y)\psi,y) = R_\rho(\xi,x) \Lambda_M(\psi,y) 
	\ . 
\labl{eq:repn-mult-prop}
Demanding that $\Omega$ acts as the identity can be formulated as
\be 
  \lim_{x \rightarrow 0} R_\rho(\xi,x) \Omega = \xi \ .
\labl{eq:repn-unit-prop}
These two equations, which have to hold for all 
$\xi \in \Vc_{-\frac18}$ and $\psi \in \Rc_0$ result in constraints on
$\rho$  and $M$. Using the two blocks  \erf{eq:mmoo-Gx} and
\erf{eq:Gtilde-sol} one finds the constraints to be, in terms of the
dual of the basis \erf{eq:H0om-basis}, and in terms of the components
$m_{11},\dots$ of the linear map $M$ in 
\erf{eq:M-general},
\be
  \rho = \big(\tfrac12 m_{11} - 2 \log 2\big) v_1^* - v_4^*
  \ , \qquad
  \big(m_{11} - 2 m_{22}\big)^2 = - 4 m_{23} m_{32} \ .
\labl{eq:m22-and-rho}
In particular, it thus follows that the OPE of $\xi$ and $\psi$ is
uniquely determined by the OPE on $(\Nr,\pm)$. Furthermore, not all
values for $M$ are consistent with the requirement that 
$\Vc_{-\frac18}$ is a space of boundary changing fields. Note also
that \erf{eq:m22-and-rho} together with $m_{11}=0$ fixes the leading
term in the OPE \erf{eq:omom-ope-via-M} completely, and one obtains
precisely \erf{eq:summary-omom-bnd-ope}.

Nonetheless, the analysis so far does not fully determine the OPE of
$\omega$ with itself beyond leading order; even after fixing the
freedom to redefine $\omega$ by setting $m_{11}=0$, we still have not 
determined the values of $m_{23}$ and $m_{32}$. While it seems likely 
that these two constants can be fixed by other factorisation
considerations, for the present paper we content ourselves with
knowing the leading order. 

\medskip

In the language of algebras and representations,  
the conditions \erf{eq:repn-mult-prop} and \erf{eq:repn-unit-prop}
define a (right-)module of the unital algebra $A_M$ defined in
section \ref{app:alg-assoc}. The conditions \erf{eq:m22-and-rho} then 
mean that only for specific choices of $M$ is it possible to endow
the object $\Vc_{-\frac18}$ with the structure of an $A_M$ module. 
However, if it is possible, the module structure is unique.


\end{document}